\documentclass[preprint, amsmath, amssymb, aps]{revtex4-2}%
\usepackage{amssymb}
\usepackage{amsfonts}
\usepackage{amsmath}
\usepackage{graphicx}%
\usepackage{dcolumn}%
\usepackage{bm}%
\usepackage[utf8]{inputenc}
\usepackage[T1]{fontenc}
\providecommand{\U}[1]{\protect\rule{.1in}{.1in}}
\newtheorem{theorem}{Theorem}

\newtheorem{lemma}[theorem]{Lemma}

\newenvironment{proof}[1][Proof]{\noindent\textbf{#1.} }{\ \rule{0.5em}{0.5em}}
\ifx\pdfoutput\relax\let\pdfoutput=\undefined\fi
\newcount\msipdfoutput
\ifx\pdfoutput\undefined\else
\ifcase\pdfoutput\else
\ifx\paperwidth\undefined\else
\ifdim\paperheight=0pt\relax\else\pdfpageheight\paperheight\fi
\ifdim\paperwidth=0pt\relax\else\pdfpagewidth\paperwidth\fi
\fi\fi\fi
\begin{document}
\title[Constrained Dynamics]{Constrained Dynamics: Generalized Lie
  Symmetries, Singular Lagrangians, and the Passage to Hamiltonian
  Mechanics} 
\author{Achilles D. Speliotopoulos}
\affiliation{Department of Physics, University of California, Berkeley, CA 94720 USA}
\altaffiliation[Also at ]{Division of Physical, Biological and Health Science, Diablo Valley College, Pleasant Hill, CA 94523, USA}
\email{ads@berkeley.edu}
\date{\today}

\nopagebreak

\begin{abstract}
Guided by the symmetries of the Euler-Lagrange equations of motion, a
study of the constrained dynamics of singular Lagrangians is
presented. We find that these equations of motion admit a generalized
Lie symmetry, and on the Lagrangian phase space the generators of this
symmetry lie in the kernel of the Lagrangian two-form. Solutions of
the energy equation\textemdash called second-order, Euler-Lagrange
vector fields (SOELVFs)\textemdash with integral flows that have this
symmetry are determined. Importantly, while second-order, Lagrangian
vector fields are not such a solution, it is always possible to
construct from them a SOELVF that is. We find that all SOELVFs are
projectable to the Hamiltonian phase space, as are all the dynamical
structures in the Lagrangian phase space needed for their
evolution. In particular, the primary Hamiltonian constraints can be
constructed from vectors that lie in the kernel of the Lagrangian two-form, and
with this construction, we show that the Lagrangian constraint
algorithm for the SOELVF is equivalent to the stability analysis of
the total Hamiltonian. Importantly, the end result of this stability
analysis gives a Hamiltonian vector field that is the projection of
the SOELVF obtained from the Lagrangian constraint algorithm. The
Lagrangian and Hamiltonian formulations of mechanics for singular
Lagrangians are in this way equivalent.   
\end{abstract}

\maketitle

\section{Introduction\label{&Intro}}

The Lagrangian phase space formulation of mechanics \cite{Abr1978,
  Car1990a, Gra2004, Car1988a}, with its roots  
in differential geometry, provides an especially fruitful
framework with which to analyze dynamical systems of singular
Lagrangians $L$. Instead of trajectories $q(t) = (q^1(t), 
\dots, q^D(t))$ on a $D$-dimensional configuration space $\mathbb{Q}$ that are
solutions of the Euler-Lagrange equations of motion, trajectories in
the Lagrangian phase space formulation  
\begin{equation}
  \mathfrak{u}(t) = (q^1(t), \dots, q^D(t), v^1(t), \dots, v^D(t)),
\end{equation}
are on a $2D$-dimensional \textbf{Lagrangian (or velocity)
  phase space} $\mathbb{P}_L=\mathbf{T}\mathbb{Q}$ embodied with a
Lagrangian two-form $\mathbf{\Omega}_L$. They are
determined by vector fields
$\mathbf{X}_E\in\mathbf{T}_{\mathfrak{u}}\mathbb{P}_L$ 
that are solutions of the \textbf{energy equation}
\begin{equation}
  i_{\mathbf{X}_E}\mathbf{\Omega}_L=\bm{d}E,
  \label{EnergyE}
\end{equation}
with $E$ being the energy of the system. For regular Lagrangians,
$\mathbf{\Omega}_L$ is symplectic. The solution to
Eq.~$(\ref{EnergyE})$ is unique, and is a second-order, Lagrangian vector
field (SOLVF)\cite{Abr1978} (also called a second-order
  dynamical equation in the literature). For singular
Lagrangians, on the other hand, $\mathbf{\Omega}_L$ is presymplectic
\cite{foot0}. The solution to Eq.~$(\ref{EnergyE})$ is not unique,
need not be a SOLVF, nor need it even exist
\cite{Got1979}. Nevertheless, with few exceptions in the literature
\cite{deL1995}, focus has been placed on solutions of
Eq.~$(\ref{EnergyE})$ that are SOLVFs. This 
is done for physical reasons: the condition $\dot{q}^a=v^a$, for
$a=1, \dots, D$, immediately follows for trajectories
determined by such fields. This focus on SOLVFs has consequences,
however.  

The presence of a singular Lagrangian often predicates the existence
of a Lagrangian constraint submanifold of $\mathbb{P}_L$, and for
solutions of Eq.~$(\ref{EnergyE})$ to exist, trajectories of
these dynamical systems must lie on this
submanifold. Algorithms\textemdash called a constraint algorithm or a 
stability condition\textemdash for constructing 
such solutions have been developed \cite{Got1978, Got1979, Got1980,
  Car1993, Mun1992, Pon1988, deL2001, Cen2014}. However, irrespective
of the one used, the end result of these algorithms is a
vector field that has a number of troubling attributes. First, this
vector field need not be a SOLVF, even though physical arguments were
used to restrict the starting point of these algorithms to such 
fields. This is called the second order problem, first noted within  
a different context by K\"unzle \cite{Kun1969}, and emphasized by Gotay and
Nester \cite{Got1980}. Currently, it is known that requiring the end
result of these algorithms be a SOLVF is very restrictive, and 
additional conditions may need to be imposed \cite{Car1987a}. Second,
the fibre derivative (Legendre transform) $\mathcal{L}$ for 
singular Lagrangians is singular, and thus the rank of the Hessian of
$L$ is not maximal. Because of this the passage from
Lagrangian to Hamiltonian mechanics is problematic. The ability to map
dynamical structures from the Lagrangian to the Hamiltonian phase
space has long been studied for SOLVFs 
\cite{Got1978, Bat1987a, Bat1987b, Car1987a, Car1987b, Car1988a,
  Pon1988, Gra1989, Mun1992, Gra1992a, Gra1992b, Pon1999, Gra2001}. It
is found that a general SOLVF\textemdash even after the application of
the constraint algorithm, and even under the weak projectability
condition \cite{Mun1992}\textemdash is not projectable \cite{Car1987a,
  Mun1992}. (Examples of systems for which a SOLVF is projectable,
and for which it is not are given in \cite{Mun1992}.) One
immediate consequence of this non-projectability is
dynamical systems for which the Hamiltonian flow field determined
through constrained Hamiltonian mechanics as described in 
\cite{Dir1950, Hen1992} (see \cite{Mun1989, Lus2018} for more
modern approaches)\textemdash even after its restriction to the primary 
constraint submanifold\textemdash need
have little relation to the SOLVF obtained from the Lagrangian constraint
algorithm. Third, it is known that the Lagrangian constraints
obtained while a constraint algorithm is being imposed on a SOLVF also need
not be projectable \cite{Bat1986, Bat1987b, Car1987a, Car1987b,
  Car1988a, Gra1992b, Mun1992}. Combined, this means that
dynamics on the Lagrangian phase space and dynamics on the Hamiltonian
phase space can take place on two inequivalent submanifolds, be
determined by two inequivalent vector fields, resulting in two
\textit{different} families of trajectories on the configuration space
$\mathbb{Q}$ for the \textit{same} dynamical system with the \textit{same}
initial data.   

We take a different starting point in our analysis of singular
Lagrangians, one that is rooted in the generalized Lie symmetries of the
Euler-Lagrange equations of motion. The analysis is guided by the
observation that if a symmetry of the dynamical system has been
determined through Lagrangian mechanics, it must be present in
the Lagrangian and Hamiltonian phase space descriptions of
motion as well. We emphasize, however, that while these symmetries play an important
role, this role is nevertheless supportive. One of the main goals of
this work is the construction of algebraic-geometric structures within
differential geometry that will then be used to implement these
symmetries; to characterize the relevant geometry of the Lagrangian
phase space; to determine the structures needed to decribe dynamics on
this phase space; and to show the equivalence of the
Lagrangian and Hamiltonian phase space formulation of mechanics. In
doing so, we are led to construct the
\textbf{second-order, Euler-Lagrange vector field} (SOELVF). These fields 
avoid the second-order problem, are projectable to the Hamiltonian
phase space, and lie on Lagrangian constraint submanifolds that also are
projectable. Importantly, the projection of the SOELVF is the
Hamiltonian flow field of the total Hamiltonian obtained from
constrained Hamiltonian mechanics. (We follow the terminology in
\cite{Hen1992}, and call the result of augmenting the canonical
Hamiltonian with the primary constraint functions the \textbf{total
  Hamiltonian}.)  

The generalized Lie symmetry \cite{Olv1993} of the Euler-Lagrange
equations of motion is generated by second-order prolongation 
vectors in the tangent space $\mathbf{T}\mathbb{M}^{(2)}$ of the
second-order jet space $\mathbb{M}^{(2)}$. This symmetry is reflected
in the Lagrangian phase space description of motion, and the projection of 
$\mathbf{T}\mathbb{M}^{(2)}$ to $\mathbf{T}\mathbb{P}_L$ maps these
prolongation vectors into the kernel of
$\mathbf{\Omega}_L$. Surprisingly, it is not the vertical vector
fields of the kernel that generates this symmetry, as 
may have been expected. Also surprisingly, the corresponding symmetry
group $\hbox{Gr}_{\mathcal{S}\hbox{ym}}$ is not a symmetry
group for SOLVFs; action on a SOLVF by
$\hbox{Gr}_{\mathcal{S}\hbox{ym}}$ results in a vector field  
that is no longer a SOLVF, nor need it even be a solution of
Eq.~$(\ref{EnergyE})$. It is, however, always possible to construct
from a SOLVF vector fields that do have
$\hbox{Gr}_{\mathcal{S}\hbox{ym}}$ as a symmetry group, and are  
solutions to Eq.~(\ref{EnergyE}). These vector fields are the SOELVFs,
and they resolve the issues listed above for the SOLVF.  

That a SOELVF is projectable is a natural consequence of having  
$\hbox{Gr}_{\mathcal{S}\hbox{ym}}$ as a symmetry
group. Moreover, all the Lagrangian 
constraints\textemdash both those due to the energy equation and those
introduced through the application of the constraint algorithm to the
SOELVF\textemdash are projectable as well. We find also that there is
a choice of a basis for the kernel of $\mathbf{\Omega}_L$ that is
projectable, and that the primary Hamiltonian constraints can be
constructed from their image. Indeed, all the dynamical structures
needed to describe evolution on the Lagrangian phase space are
projectable, and their image corresponds to the dynamical structures
needed to describe evolution on the Hamiltonian phase space obtained
through constrained Hamiltonian mechanics \cite{Dir1950, Hen1992,
  Mar1983, Mun1989, Gra1991, Bat2013, Cen2014, Lus2018}. In this way
the Lagrangian and Hamiltonian fomulations of mechanics are equivalent 
even for singular Lagrangians.

Analysis of the symmetries of Lagrangian systems (both regular and
singular) have been done before. However, such analyses have
been focused on time-dependent Lagrangians (and in particular their Noether
symmetries) \cite{Pri1983, Pri1985, 
Cra1983, Car1991, Car1988b, Car1992, Car1993, Car2003}; on systems of
first-order evolution equations on either the Lagrangian or
Hamiltonian phase space \cite{Car1990b, Mar1992, Gra2002, Gra2005,
  Pop2017}; or on general solutions of Eq.~$(\ref{EnergyE})$ 
\cite{deL1995} (see also \cite{Dim2016} for an analysis of particle
motion on curved spacetimes). Importantly, the great majority of these
analyses have been done using first-order prolongations on
first-order jet bundles with a focus on the Lie symmetries of first-order
evolution equations. Our interest is in the symmetries of the
Euler-Lagrange equations of motion\textemdash a system of  
second-order differential equations\textemdash that come from singular
Lagrangians. This naturally leads us to consider generalized Lie
symmetries and second-order prolongations. Such symmetry analysis of
the Euler-Lagrange equations of motion has not been done
before. (Although the framework for $k^{\hbox{\textit{th}}}$-order
prolongations on $k^{\hbox{\textit{th}}}$-order jet bundles have been
introduced before \cite{Car1993, Car2003, Pop2009,
  Pop2011}, they have not been applied to the Euler-Lagrange equations
of motion.)

In this paper we only consider
autonomous Lagrangians for which the rank of $\mathbf{\Omega}_L$ is
constant on $\mathbb{P}_L$. We also require such Lagrangians to have
a fibre derivative that is a submersive map of $\mathbb{P}_L$ to the
Hamiltonian phase space $\mathbb{P}_C$; the rank of the Hessian of such
Lagrangians is necessarily constant on $\mathbb{P}_L$. In addition, the
preimage of $(q,p)=\mathcal{L}(\mathfrak{u}),
\mathfrak{u}\in\mathbb{P}_L$, must be a connected submanifold of
$\mathbb{P}_L$ (see also \cite{Bat2013} where this condition is
relaxed). These Lagrangians are called \textbf{almost regular}
Lagrangians in the literature \cite{Got1978, Car1987a, Car1988a,
  Car1990a, Gra2001}, and we also use this terminology.  

Some of the results on the equivalence of singular Lagrangian and
Hamiltonian mechanics presented in this paper have been presented
elsewhere. However, the approaches used previously often rely on
such dynamical structures from constrained Hamiltonian mechanics as 
the primary Hamiltonian constraints; pullbacks of their
derivatives to the Lagrangian phase space are used to construct 
such mappings as the time-evolution operator $K$ \cite{Car1988b,
  Bat1987a, Car1987b, Pon1988, Gra1989, Gra1992b}, for example. We
take a different approach, one that starts with the Lagrangian phase
space formulation, and, with restrictions imposed by
$\hbox{Gr}_{\mathcal{S}\hbox{ym}}$, is one which shows
that the dynamical structures on the Hamiltonian phase
space necessary to describe the dynamical system can be obtained
directly from those on the Lagrangian phase space.  

The importance of establishing the equivalence between the Lagrangian
and Hamiltonian formulations of mechanics for singular Lagrangians can
been seen from the starting point of any physically relevant system:
the action, and through it, its Lagrangian. For systems with local
gauge or diffeomorphism symmetry, the Lagrangian is singular, and
dynamics have traditionally been analysed using Hamiltonian constraints
\cite{Hen1992}. This is done by first using the fibre
derivative to construction the Hamiltonian from the Lagrangian, and then
using Hamiltonian stability analysis to determine both the total
Hamiltonian and the Hamiltonian constraint surfaces. However, as the
Lagrangian was the starting point, and as the fibre derivative is not invertable
for such Lagrangians, a natural question to ask is whether the dynamics
described by the total Hamiltonian has any relation to the original
dynamics given by the Lagrangian. With the equivalence between the
Lagrangian and the Hamiltonian phase space formulations of dynamics
demonstrated, we have shown that they are. This equivalence is even
more important for the path integral formulation of quantum mechanics
and quantum field theory. Both are based on the action, and
integration is over paths on the configuration space. For systems with
local gauge or diffeomorphism symmetries these integrals must be
restricted, which for non-abelian gauge theories leads to the use of BRST
symmetries. These symmetries have traditionally been constructed using
the Hamiltonian and Hamiltonian constraint analysis (see
\cite{Hen1992}).   

Although we freely use the tools and language of differential
geometry, we are aware that interest in constrained dynamics is often
due to its application to quantum field theories. In these
applications, the ability to calculate and determine symmetries is
paramount. To ensure that the tools and methodologies given in this paper
can be so applied to the analysis of quantum field-theoretic
systems, we have also written a number of the expressions given in
this paper in terms of local coordinates using a notation that is both
familiar and useful for calculations. In particular, the general
solution to $i_{\mathbf{K}}\bm{\Omega}_L=0$ given in \textbf{Section
  \ref{Prop}} is given in terms of local coordinates as are the 
construction of second-order Lagrangian and Euler-Lagrangian vector
fields.  
      
The rest of the paper is arranged as follows. In \textbf{Section
\ref{&L-Sym}} we show that the Euler-Lagrange equations of motion have
a generalized Lie symmetry, and determine the existence conditions for
the generators of this symmetry. In
\textbf{Section \ref{&LagnPhaseSp}} the vectors that lie in the kernel
of $\mathbf{\Omega}_L$ are found, and the role they play in
generating $\hbox{Gr}_{\mathcal{S}\hbox{ym}}$ is
determined. Physically relevant solutions of the energy equation are
characterized, and the SOELVF is defined and constructed. First-order
Lagrangian constraints are also constructed, and a constraint
algorithm for SOELVFs is presented. In 
\textbf{Section \ref{Passage}} focus is on the passage from the
Lagrangian to the Hamiltonian phase space. The projectability of
functions on $\mathbb{P}_L$ is reviewed, and a new result on the
projectability of vector fields in
$\mathbf{T}\mathbb{P}_L$ is presented. The dynamical 
structures needed to describe evolution with SOELVFs are shown to be
projectable, and the primary Hamiltonian constraints are
constructed. The equivalence of the constraint algorithm presented in
\textbf{Section \ref{&LagnPhaseSp}} with the usual Hamiltonian
stability analysis is shown. In \textbf{Section \ref{Exam}} application 
of the analysis given here to three different dynamical systems with
singular Lagrangians is presented. Concluding remarks can be found in
\textbf{Section \ref{Conc}}, with the crucial role that
the vertical vector fields in the kernel of $\mathbf{\Omega}_L$ play
summarized.    

\section{Generalized Lie symmetries and Lagrangian mechanics\label{&L-Sym}}

We begin with Lagrangian mechanics, and an analysis of the
generalized Lie symmetry \cite{Olv1993} of the Euler-Lagrange
equations of motion. The existence conditions for the generators of
this symmetry will be established.  

The Euler-Lagrange equations of motion are the system of $D$
second-order differential equations
\begin{equation}
M_{ab}\ddot{q}^{b}=-\frac{\partial E}{\partial q^{a}}-F_{ab}\dot{q}^{b},
\label{2ndEL1}%
\end{equation}
where 
\begin{equation}
E\left(  q,\dot{q}\right)  :=\dot{q}^{b}\frac{\partial L\left(  q,\dot
{q}\right)  }{\partial\dot{q}^{b}}-L\left(  q,\dot{q}\right)  , \label{Leng}%
\end{equation}%
is the energy, while
\begin{equation}
M_{ab}\left(  q,\dot{q}\right)  :=\frac{\partial^{2}L\left(  q,\dot{q}\right)
}{\partial\dot{q}^{a}\partial\dot{q}^{b}}, \quad \hbox{and }\quad 
F_{ab}\left(  q,\dot{q}\right)  := \frac{\partial^{2}L\left(
  q,\dot{q}\right)  }{\partial\dot{q}^{a}\partial q^{b}} -
\frac{\partial^{2}L\left(  q,\dot{q}\right)}{\partial\dot{q}^{b}\partial 
  q^{a}}.
\end{equation}
Here, Einstein's summation convention is used.

For almost regular Lagrangians the rank of the Hessian
$M_{ab}(\mathfrak{u})$, where 
$\mathfrak{u}=\left(q,\dot{q}\right)$ \cite{foot3}, is constant on
$\mathbb{P}_L$. However, as the rank of $M_{ab}\left( 
  \mathfrak{u}\right) =D-N_{0}$, with $N_0=\hbox{dim
}\left(\hbox{ker }M_{ab}(\mathfrak{u})\right)$, this rank is not maximal, and
thus Eq.~$(\ref{2ndEL1})$ cannot be solved for a unique
$\ddot{q}$. Instead, a chosen set of initial data
$\mathfrak{u}_{0}=\left(  q_{0},\dot{q}_{0}\right)$ given at
$t=t_{0}$ determines a family of solutions to Eq.~$(\ref{2ndEL1})$
that evolve from the same $\mathfrak{u}_0$. These solutions
are related to one another through a generalized Lie
symmetry \cite{Olv1993}.   

Following Olver \cite{Olv1993}, we define
\begin{equation}
  \Delta_a(q,\dot{q}, \ddot{q}) := \frac{\partial E(q, \dot{q})}{\partial q^a} +
  F_{ab}(q, \dot{q})\dot{q}^b + M_{ab} \ddot{q}^b,  
\label{delta}
\end{equation}
which reduces to Eq.~(\ref{2ndEL1}) on the surfaces
$\Delta_a(q,\dot{q}, \ddot{q})=0$. The set  
\begin{equation}
\mathcal{O}\left(  \mathfrak{u}_{0}\right)  :=\big\{q\left(  t\right)
\ \backslash\ \
\Delta_a(q,\dot{q}, \ddot{q}) =0 \hbox{ with }
q\left(  t_{0}\right)
=q_{0},\ \dot{q}\left(  t_{0}\right)  =\dot{q}_{0}\big\}  ,
\label{gaugeorb1}
\end{equation}
is the family of solutions to Eq.~$(\ref{2ndEL1})$ that evolve from
$\mathfrak{u_0}$. Consider two such solutions $q^a(t)$ and
$Q^a(t)$. From Eq.~$(\ref{2ndEL1})$ there exists a 
$\mathfrak{z}^a(\mathfrak{u})\in\hbox{ker } 
M(\mathfrak{u})$ such that $\ddot{Q}^a-\ddot{q}^a =
\mathfrak{z}^a(\mathfrak{u})$. As $\mathfrak{z}^a$ depends on both $q^a$ and
$\dot{q}^a$, we are led to consider generalized Lie symmetry groups
generated by
\begin{equation}
  \mathbf{g} := \rho(\mathfrak{u})\cdot \frac{\bm{\partial}
    \>\>\>}{\bm{\partial} q}, 
\end{equation}
with a $\rho(\mathfrak{u})$ that does not depend explicitly on time. This
gives the total time derivative
\begin{equation}
  \frac{\bm{d}\>\>\>}{\bm{d}t} := \dot{q}\cdot
  \frac{\bm{\partial}\>\>\>}{\bm{\partial} q} +\ddot{q}\cdot
  \frac{\bm{\partial} \>\>\>}{\bm{\partial} \dot{q}}, 
\end{equation}
with $\dot{\rho}:=\bm{d}\rho/\bm{d}t$.

With this $\mathbf{g}$, the second-order prolongation vector,
\begin{equation}
  \hbox{\textbf{pr }}\mathbf{g} := \rho \cdot
  \frac{\bm{\partial}\>\>\>}{\bm{\partial} q} 
  + \dot{\rho} \cdot \frac{\bm{\partial}\>\>\>}{\bm{\partial} \dot{q}} +
  \ddot{\rho}\cdot \frac{\bm{\partial} \>\>\>}{\bm{\partial} \ddot{q}},
\end{equation}
on the second-order jet space $\mathbb{M}^{(2)}=\{(t, q,
\dot{q},\ddot{q})\}$ with $\hbox{\textbf{pr
}}\mathbf{g}\in\mathbf{T}\mathbb{M}^{(2)}$ can be constructed. Its
action on the Euler-Lagrange equations of motion gives  
\begin{equation}
  \hbox{\textbf{pr }}\mathbf{g}\left[\Delta_a(q,
    \dot{q},\ddot{q})\right] = -\frac{\partial 
    \ddot{q}^b}{\partial q^a}M_{bc}\rho^c +
  \frac{d\>\>\>}{dt}\left[F_{ab}\rho^b + M_{ab}\dot{\rho}^b\right],
  \label{e1}
\end{equation}
on the $\Delta_a=0$ surface. However, because the
rank of $M_{ab}(\mathfrak{u})$ is not maximal,  
the solution for $\ddot{q}$ on this surface is not unique. For
$\mathbf{g}$ to generate the same symmetry group for all the
trajectories in $\mathcal{O}(\mathfrak{u}_0)$, we must require
$\rho(\mathfrak{u})\in\hbox{ker }M_{ab}(\mathfrak{u})$. It then
follows that $\hbox{\textbf{pr g}}[\Delta_a(q,\dot{q}, \ddot{q})] =0$
if and only if (iff) $b_a = 
F_{ab}\rho^b + M_{ab}\dot{\rho}^b$ for some function
$b_a(\mathfrak{u})$ where $\dot{b}_a=0$. However, because all the
solutions in $\mathcal{O}(\mathfrak{u}_0)$ have the same initial data, 
$\rho^a(\mathfrak{u}_0)=0$ and $\dot{\rho}^a(\mathfrak{u}_0)=0$,
and thus $b_a=0$. This leads to the following new result.

\begin{lemma} \label{GS} If $\mathbf{g}$ is
  a generalized infinitesimal symmetry of $\Delta_a$, then
  $\rho^a(\mathfrak{u})\in\hbox{ker } M_{ab}(\mathfrak{u})$ and
  $\dot{\rho}^a(\mathfrak{u})$ is a solution of
    \begin{equation}
    0=F_{ab}(\mathfrak{u})\rho^b(\mathfrak{u}) +
    M_{ab}(\mathfrak{u})\dot{\rho}^a(\mathfrak{u}).
    \label{sol}
   \end{equation}
\end{lemma}
We denote by $\mathfrak{g}$ the set of all vector fields $\mathbf{g}$
that satisfy \textbf{Lemma $\mathbf{\ref{GS}}$}, and by
$\hbox{\textbf{pr }}\mathfrak{g} := \{\hbox{\textbf{pr
}}\mathbf{g}\ \backslash\ \ \mathbf{g}\in \mathfrak{g}\}$ the set of
their prolongations. This $\hbox{\textbf{pr }}\mathfrak{g}$ is involutive
\cite{Olv1993}.  

The conditions under which $\hbox{\textbf{pr }}\mathfrak{g}$
generates a generalized Lie 
symmetry group are well known \cite{Olv1993}. However, because our
Lagrangians are singular, three additional conditions must be imposed: 

\begin{enumerate}
\item{While $\rho^a=0$ and $\dot{\rho}^a=\mathfrak{z}^a$ for any
$\mathfrak{z}\in\hbox{ker }M_{ab}(\mathfrak{u})$ is a solution of
  Eq.~$(\ref{sol})$, we require that $\dot{\rho}^a =
  \bm{d}\rho^a/\bm{d}t$, and they must be removed.} 
\item{If $\dot{\rho}^a$ is a solution of Eq.~$(\ref{sol})$, then
$\dot{\rho}^a+\mathfrak{z}^a$ is a solution of Eq.~$(\ref{sol})$ as
  well. The $\dot{\rho}^a$ are not unique, and this, along with the
  first condition, leads us to generators that are constructed from
  equivalence classes of prolongations.} 
\item{For any $\mathfrak{z}^a \in\hbox{ker }M_{ab}(\mathfrak{u})$,
  Eq.~$(\ref{delta})$ 
  gives,
\begin{equation}
  0=\mathfrak{z}^a\left(\frac{\partial E}{\partial
    q^a}+F_{ab}(q,\dot{q})\dot{q}^b\right),
  \label{e}
\end{equation}
on the solution surface $\Delta_a(q,\dot{q},\ddot{q})=0$. If
Eq.~(\ref{e}) does not hold identically, it must be
imposed, leading to the well-known, first-order Lagrangian
constraints. As each $q(t)\in\mathcal{O}(\mathfrak{u}_0)$ 
must lie on this constraint submanifold, any symmetry transformation of
$q(t)$ generated by $\mathbf{pr }\>\> \mathbf{g}$ must give a path
$Q(t)$ that also lies on the constraint submanifold.} 
\end{enumerate}

Not all vectors in $\mathbf{pr}\> \mathfrak{g}$ will be generators of
the generalized Lie symmetry group for 
$\mathcal{O}(\mathfrak{u}_0)$. Determining which vectors are is best
done within the Lagrangian phase space framework. This will be done in
the next section. For now, we note the following.    

For any $\mathbf{g}\in\mathfrak{g}$, decompose
\begin{equation}
  \hbox{\textbf{pr} }\mathbf{g} = \mathbf{k} +
  \ddot{\rho}\cdot\frac{\bm{\partial} \>\>\>}{\bm{\partial}\ddot{q}}. 
\end{equation}
Then for $\hbox{\textbf{pr }}\mathbf{g}_A, \hbox{\textbf{pr
}}\mathbf{g}_B, \in \hbox{\textbf{pr }}\mathfrak{g}$, 
\begin{equation}
  [\hbox{\textbf{pr }}\mathbf{g}_A, \hbox{\textbf{pr }}\mathbf{g}_B] =
  [\mathbf{k}_A, \mathbf{k}_B] +
  \left[\hbox{\textbf{pr }}\mathbf{g}_A (\ddot{\rho}^a_B) -
    \hbox{\textbf{pr }}\mathbf{g}_B (\ddot{\rho}^a_A)
      \right]\frac{\bm{\partial}\>\>\>}{\bm{\partial} \ddot{q}^a}.
\end{equation}
Because
$\hbox{\textbf{pr }}\bm{\mathfrak{g}}$ is involutive, there
exists a \textbf{pr }$\mathbf{g}_C\in\hbox{\textbf{pr }}\mathfrak{g}$
such that $\hbox{\textbf{pr }}\mathbf{g}_C = [\hbox{\textbf{pr
  }}\mathbf{g}_A, \hbox{\textbf{pr }}\mathbf{g}_B]$. There is then a
$\mathbf{k}_C$ such that $\mathbf{k}_C=[\mathbf{k}_A,\mathbf{k}_B]$,
and the collection of vectors 
\begin{equation}
    \mathcal{K} = \Bigg\{\mathbf{k} =
    \rho(\mathfrak{u})\cdot\frac{\bm{\partial} \>\>\>}{\bm{\partial}
      q} + \dot{\rho}(\mathfrak{u})\cdot \frac{\bm{\partial}
      \>\>\>}{\bm{\partial}\dot{q}} \ \backslash\  
    \rho^a(\mathfrak{u})\in\hbox{ker }M_{ab}(\mathfrak{u}),
    0 = F_{ab}(\mathbf{u})\rho^b(\mathbf{u})
    + M_{ab}(\mathbf{u})\dot{\rho}^b(\mathbf{u})\Bigg\},
\end{equation}
is involutive. Importantly, dim $\mathbf{pr \>\>}\mathfrak{g} =
\hbox{dim }\mathcal{K}=2N_0$.    

\section{The Lagrangian phase space\label{&LagnPhaseSp}}

In this section we determine the generators of the
generalized Lie symmetry found in \textbf{Section
  \ref{&L-Sym}}. This is done on the Lagrangian phase space using the
tools of differential geometry. These generators are then used to
determine the physically relevant solutions of the energy equation,
and with them, the constraint algorithm and the Lagrangian constraint
submanifold. Much of the content in \textbf{Sections \ref{Pass1}} to
\textbf{\ref{LCon}} have been established elsewhere. They are gathered
here for clarity and coherence of argument, and to establish notation
and terminology. On the other hand, much of the construction and the
results presented in \textbf{Sections \ref{Sym}} to
\textbf{\ref{&StabC}} are new.  

\subsection{Passage from Lagrangian mechanics to the Lagrangian phase
  space \label{Pass1}}

To treat singular Lagrangian dynamics using the methods of
differential geometry, trajectories $t\to q(t)\in\mathbb{Q}$ are
replaced by integral flows $t\to\mathfrak{u}(t)\in\mathbb{P}_L$
\cite{Abr1978}, which for the initial data $\mathfrak{u}_0=(q_0,
v_0)$ are given by solutions to  
\begin{equation}
\frac{d\mathfrak{u}}{dt}:=\mathbf{X}  (\mathfrak{u}),
\end{equation}
where $\mathbf{X}$ is a smooth vector field in the tangent space
$\mathbf{T}\mathbb{P}_L$. As $\mathbb{P}_L= \mathbf{T}\mathbb{Q}$, we have 
the bundle projections $\tau_{\mathbb{Q}}:\mathbf{T}\mathbb{Q}\to
\mathbb{Q}$ and 
$\tau_{\mathbf{T}\mathbb{Q}}:\mathbf{T}(\mathbf{T}\mathbb{Q})
\to\mathbf{T}\mathbb{Q}$ with $\tau_{\mathbb{Q}}\circ
\tau_{\mathbf{T}\mathbb{Q}}:\mathbf{T}(\mathbf{T}\mathbb{Q})  
\to\mathbb{Q}$ (see \cite{Got1979} and \cite{Abr1978}). It is also
possible to construct $\mathbf{T}\tau_{\mathbb{Q}}$, the prolongation of 
$\tau_{\mathbb{Q}}$ to $\mathbf{T}(\mathbf{T}\mathbb{Q})$, which is
a second projection map
$\mathbf{T}\tau_{\mathbb{Q}}: \mathbf{T}(\mathbf{T}\mathbb{Q}) 
\to\mathbf{T}\mathbb{Q}$ which is determined by requiring that
$\tau_{\mathbb{Q}}\circ \tau_{\mathbf{T}\mathbb{Q}}$ and
$\tau_{\mathbb{Q}}\circ \mathbf{T}\tau_{\mathbb{Q}}$ map a point in
$\mathbf{T}(\mathbf{T}\mathbb{Q})$ to the same point in
  $\mathbb{Q}$. The \textbf{vertical subbundle}
$[\mathbf{T}\mathbb{P}_L]^v$ of $\mathbf{T}(\mathbf{T}\mathbb{Q})$
is defined by $[\mathbf{T}\mathbb{P}_L]^v = \hbox{ker
}\mathbf{T}\tau_{\mathbb{Q}}$ \cite{Got1979}; a vector $\mathbf{X}^v\in
[\mathbf{T}_{\mathfrak{u}}\mathbb{P}_L]^v$ above a point
$\mathfrak{u}\in\mathbb{P}_L$ is called a \textbf{vertical vector
  field}. The \textbf{horizontal subbundle} 
$[\mathbf{T}\mathbb{P}_L]^q$ of
$\mathbf{T}(\mathbf{T}\mathbb{Q})$ is defined by
$[\mathbf{T}\mathbb{P}_L]^q = \hbox{Image
}\mathbf{T}\tau_{\mathbb{Q}}$; a vector
$\mathbf{X}^q\in[\mathbf{T}_{\mathfrak{u}}\mathbb{P}_L]^q$ is called a 
\textbf{horizontal vector field}. Each  
$\mathbf{X}\in\mathbf{T}_{\mathfrak{u}}\mathbb{P}_L$ can be
expressed as $\mathbf{X} = \mathbf{X}^q + \mathbf{X}^v$ with
$\mathbf{X}^q \in
\left[\mathbf{T}_{\mathfrak{u}}\mathbb{P}_L\right]^q$ and
$\mathbf{X}^v \in
\left[\mathbf{T}_{\mathfrak{u}}\mathbb{P}_L\right]^v$, which in terms
of local coordinates are,  
\begin{equation}
\mathbf{X}^{q}
:=X^{qa}\frac{\bm{\partial}\>\>\>}{\bm{\partial}q^{a}} ,\quad
\hbox{and}\quad\mathbf{X}^{v} :=X^{va}
\frac{\bm{\partial}\>\>\>}{\bm{\partial}v^{a}}.  \label{VecFld1}%
\end{equation}
In particular, a second order Lagrangian vector field $\mathbf{X}_L$
is the solution of Eq.~$(\ref{EnergyE})$ such that
$\mathbf{T}\tau_{\mathbb{Q}}\circ\mathbf{X}_L$ is the identiy on
$\mathbf{T}\mathbb{Q}$ (see \cite{Abr1978}). In terms of local
coordinates
\begin{equation}
  \mathbf{X}_L = v^a \frac{\bm{\partial}\>\>\>}{\bm{\partial} q^a} +
  X^{va} \frac{\bm{\partial}\>\>\>}{\bm{\partial}v^a},
\end{equation}
where for singular Lagrangians $X^{va}$ is not unique.

For a one-form $\bm{\alpha}\in
\mathbf{T}^{*}_{\mathfrak{u}}\mathbb{P}_L$, where
$\mathbf{T}^{*}\mathbb{P}_L$ is the cotangent space of
one-forms on $\mathbb{P}_L$, and a vector
field $\mathbf{X}\in\mathbf{T}_{\mathfrak{u}}\mathbb{P}_L$, we define the
dual prolongation map $\mathbf{T}^{*}\tau_{\mathbb{Q}}$ by
\begin{equation}
  \langle \bm{\alpha}\vert
  \mathbf{T}\tau_{\mathbb{Q}}\mathbf{X}\rangle = \langle
  \mathbf{T}^{*}\tau_{\mathbb{Q}} \bm{\alpha}\vert
  \mathbf{X}\rangle.
\end{equation}
Here, we adapt Dirac's bra
and ket notation to
denote the action of a $k$-form \textbf{$\bm{\omega}$}$\left(x\right)  $ by%
\begin{equation}
\bm{\omega}\left(  x\right)  :\mathbf{Y}_{1}\otimes\cdots\otimes
\mathbf{Y}_{k}\rightarrow\left\langle \left.  \bm{\omega}\left(  x\right)
\right\vert \mathbf{Y}_{1}\otimes\cdots\otimes\mathbf{Y}_{k}\right\rangle
\in\mathbb{R}, \label{kfrmg}%
\end{equation}
where 
$\mathbf{Y}_{j}\in\mathbf{T}_{\mathfrak{u}}\mathbb{P}_L$. The
$k$-form bundle is $\mathbf{\Lambda}^{k}\left(\mathbb{P}_L\right)$,
while the exterior algebra of forms is denoted by
$\mathbf{\Lambda}(\mathbb{P}_L)$. Then the \textbf{vertical one-form
  subbundle} $[\mathbf{T}^{*}\mathbb{P}_L]^v$ of
$\mathbf{T}^{*}\mathbb{P}_L$ is defined by
$[\mathbf{T}^{*}\mathbb{P}_L]^v :=\hbox{ker }
\mathbf{T}^{*}\tau_{\mathbb{Q}}$; a one-form
$\bm{\alpha}_v\in [\mathbf{T}^{*}_{\mathfrak{u}}\mathbb{P}_L]^v$ is called a
\textbf{vertical one-form}. The \textbf{horizontal
one-form subbundle} $[\mathbf{T}^{*}\mathbb{P}_L]^q$ of
$\mathbf{T}^{*}\mathbb{P}_L$ is defined by
$[\mathbf{T}^{*}\mathbb{P}_L]^q=\hbox{Image }
\mathbf{T}^{*}\mathbb{P}_L$; a one-form
$\bm{\alpha}_q\in[\mathbf{T}^{*}_{\mathfrak{u}}\mathbb{Q}]^q$ is called a
\textbf{horizontal one-form}. Each
one-form $\bm{\varphi}\in \mathbf{T}^{*}_{\mathfrak{u}}\mathbb{P}_L$
  can be expressed as $\bm{\varphi}=\bm{\varphi}_{q}
+\bm{\varphi}_{v}$ with $\bm{\varphi}_{q} \in
\left[\mathbf{T}_{\mathfrak{u}}^{*}\mathbb{P}_L\right]^q$ and
$\bm{\varphi}_{v} \in
\left[\mathbf{T}_{\mathfrak{u}}^{*}\mathbb{P}_L\right]^q$. In terms
of local coordinates 
$\bm{\varphi}_{q} :=\varphi_{qa} \ \mathbf{d}q^{a}$ and 
$\bm{\varphi}_{v} :=\varphi_{va} \mathbf{d}v^{a}$. 

There are a number approaches \cite{Abr1978} used in the literature to
obtain the \textbf{Lagrange two-form} $\bm{\Omega}_L$. We follow
\cite{Got1979, Got1980}, and define $\bm{\Omega}_L = -\mathbf{d}\mathbf{d}_JL$,
where $\mathbf{d}_J$ is the vertical derivative (see
\cite{Got1979}). This two-form can be expressed as
$\mathbf{\Omega}_{L}:=\mathbf{\Omega}_{F}+\mathbf{\Omega}_{M}$ where
\begin{equation}
\bm{\Omega}_F(\mathbf{X},\mathbf{Y}) :=
\bm{\Omega}_L(\mathbf{T}\tau_{\mathbb{Q}}\mathbf{X},
\mathbf{T}\tau_{\mathbb{Q}}\mathbf{Y}),
\end{equation}
for all $\mathbf{X},
\mathbf{Y}\in\mathbf{T}_{\mathfrak{u}}\mathbb{P}_L$; this is a
horizontal two-form of $\mathbf{\Omega}_L$. Then
$\bm{\Omega}_M(\mathbf{X},\mathbf{Y})=\bm{\Omega}_L(\mathbf{X},\mathbf{Y})- 
\bm{\Omega}_F(\mathbf{X},\mathbf{Y})$; this is a mixed two-form of
$\mathbf{\Omega}_L$. In terms of local coordinates,  
\begin{equation}
\mathbf{\Omega}_{L}=-\bm{d{\theta}}_{L},\quad \hbox{where
}\bm{\theta}_{L}=\frac{\partial L}{\partial v^{a}%
}\mathbf{d}q^{a},
\label{OmLxct}%
\end{equation}
while
\begin{equation}
\mathbf{\Omega
}_{F}:=\frac{1}{2}F_{ab}\mathbf{d}q^{a}\wedge\mathbf{d}q^{b}%
,\ \hbox{and} \ \mathbf{\Omega}_{M}:=M_{ab}\mathbf{d}q^{a}\wedge\mathbf{d}v^{b}.
\label{OmG}%
\end{equation}

If $\mathfrak{u}(t)$ is to describe the evolution of the dynamical
system given by $L$, then the vector field $\mathbf{X}$ must be chosen
so that its integral flows on $\mathbb{P}_{L}$ faithfully represent
their trajectories on $\mathbb{Q}$. For 
regular Lagrangians this is guaranteed by setting $\mathbf{X}=\mathbf{X}_{L}$,
and is a unique solution of the energy equation
Eq.~$(\ref{EnergyE})$ \cite{Abr1978}.  

We adopt the general assumption that even for singular Lagrangians
there are solutions of the energy equation that faithfully represent
trajectories on $\mathbb{Q}$. For general singular 
Lagrangians neither the existence nor the uniqueness of solutions
to Eq.~(\ref{EnergyE}) is assured. For almost regular Lagrangians,
however, a number of general results are available. These results
depend on the structure of the family of solutions evolving from
$\mathfrak{u}_{0}$, and in this the kernel of $\mathbf{\Omega}_L$,
\begin{equation}
\ker\>  \mathbf{\Omega}_{L}\left(  \mathfrak{u}\right) 
:=\left\{  \mathbf{K}\in\mathbf{T}_{\mathfrak{u}}\mathbb{P}_{L}%
\ \backslash\ \ i_{\mathbf{K}}\mathbf{\Omega}_{L}=0\right\},
\label{kerOmL66}%
\end{equation}
plays a defining role.

\subsection{Properties of ker $\mathbf{\Omega}_L(\mathfrak{u})$ \label{Prop}}

In this section we characterize the structure of ker
$\mathbf{\Omega}_L(\mathfrak{u})$, and determine the vectors that
lie in it. 

The two-form $\mathbf{\Omega}_L$ gives the lowering map
$\mathbf{\Omega}_L^{\flat}:\mathbf{T}_{\mathfrak{u}}\mathbb{P}_{L}\rightarrow 
\mathbf{T}_{\mathfrak{u}}^{\ast}\mathbb{P}_{L}$, with
$\Omega_L^{\flat}\mathbf{X}:=i_{\mathbf{X}}\mathbf{\Omega}_L$.
As $\Omega_{L}^{\flat}=\Omega_{F}^{\flat}+\Omega_{M}^{\flat}$,
the action of $\Omega_L^\flat$ on a vector
$\mathbf{X}$ is given by $\Omega_{F}^{\flat}:
\mathbf{X}\in\mathbf{T}_{\mathfrak{u}}\mathbb{P}_L \to
\left[\mathbf{T}^{*}_{\mathfrak{u}}\mathbb{P}_L\right]^q$, 
and, since $\mathbf{\Omega}_M$ is a mixed two-form, by
$\Omega_{M}^{\flat}=\Omega_{M}^{v\flat}+\Omega_{M}^{q\flat}$, where 
$\Omega_{M}^{q\flat}:\mathbf{X}\in\mathbf{T}_{\mathfrak{u}}\mathbb{P}_L 
\to\left[\mathbf{T}^{*}_{\mathfrak{u}}\mathbb{P}_L\right]^q$ and
$\Omega_{M}^{v\flat}:
\mathbf{X}\in\mathbf{T}_{\mathfrak{u}}\mathbb{P}_L \to
\left[\mathbf{T}^{*}_{\mathfrak{u}}\mathbb{P}_L\right]^v$. In terms of local
coordinates, 
$\Omega_{F}^{\flat}\mathbf{X} = F_{ab}X^{qa}\bm{d}q^b$, 
$\Omega_{M}^{q\flat}\mathbf{X}= -M_{ab}X^{va}{}\mathbf{d}q^{b}$ and
$\Omega_{M}^{v\flat}\mathbf{X}= M_{ab}X^{qa}\mathbf{d}v^{b}$.

For almost regular Lagrangians $\ker
  \Omega_{M}^{v\flat} = \mathcal{C}\oplus
  \left[ \mathbf{T}_{\mathfrak{u}} \mathbb{P}_{L}\right]  ^{v}$ 
  while $\ker  \Omega_{M}^{q\flat}  =\left[
    \mathbf{T}_{\mathfrak{u}}\mathbb{P}_{L}\right]  ^{q}\oplus  
  \mathcal{G}$, where
  \begin{equation}
    \mathcal{C}=\left\{\mathbf{C}\in[\mathbf{T}_q\mathbb{P}_L]^q
      \ \backslash \ i_{\mathbf{C}}\bm{\Omega}_M =0\right\},
  \end{equation}
  while
  \begin{equation}
    \mathcal{G}\subset\left\{\mathbf{G}\in
    [\mathbf{T}_q\mathbb{P}_L]^v \ \backslash
    \ i_{\mathbf{G}}\bm{\Omega}_M =0\right\}.
  \end{equation}
Moreover, because the rank of $M_{ab}(\mathfrak{u})$ is constant on
$\mathbb{P}_L$ there exists a basis,  
\begin{equation}
\Big\{
\bm{\mathfrak{z}}_{\left(  n\right)  }\left(  \mathfrak{u}\right) =\left(  \mathfrak{z}%
_{\left(  n\right)  }^{1}\left(  \mathfrak{u}\right)  ,\ldots,\mathfrak{z}%
_{\left(  n\right)  }^{D}\left(  \mathfrak{u}\right)  \right)  \backslash
\ M_{ab}\left(  \mathfrak{u}\right)  \mathfrak{z}_{\left(  n\right)  }%
^{b}\left(  \mathfrak{u}\right)  =0,
n=1,\ldots,N_{0} \Big\}  ,\label{bnullM}
\end{equation}
for $\ker  M_{ab}\left(  \mathfrak{u}\right) $ at each
$\mathfrak{u}\in\mathbb{P}_{L}$. This in turn gives the bases
\begin{equation}
  \mathcal{C}= \hbox{span }
  \left\{\mathbf{U}^q_{(n)}=\bm{\mathfrak{z}}_{(n)}\cdot
    \frac{\bm{\partial}\>\>\>}{\bm{\partial} q}, n=1,
    \dots, N_0\right\}, \>
  \mathcal{G}= \hbox{span }
  \left\{\mathbf{U}^v_{(n)}=\bm{\mathfrak{z}}_{(n)}\cdot
    \frac{\bm{\partial}\>\>\>}{\bm{\partial} v}, n=1,
    \dots, N_0\right\},
    \label{basesM}
\end{equation}
for $\mathcal{C}$ and $\mathcal{G}$. It is well known
\cite{Car1990a} that for almost regular Lagrangians
$\mathcal{G}$ is involutive. Furthermore, when the rank 
of $\mathbf{\Omega}_L$ is constant on $\mathbb{P}_L$, ker
$\mathbf{\Omega}_L(\mathfrak{u})$ is involutive as well.

Corresponding to $\mathbf{U}^q_{(n)}$ and $\mathbf{U}^v_{(n)}$ we have
the one-forms  
$\bm{\Theta}^{(n)}_q$ and $\bm{\Theta}^{(n)}_v$
where $\langle\bm{\Theta}^{(n)}_q\vert
\mathbf{U}^q_{(m)} \rangle= \delta_m^n$ and
$\langle\bm{\Theta}^{(n)}_v\vert \mathbf{U}^v_{(m)} \rangle=
\delta_m^n$. Then $\left[  \mathbf{T}%
_{\mathfrak{u}}\mathbb{P}_{L}\right]  ^{q}=%
\mathcal{C}%
\oplus%
\mathcal{C}%
_{\perp}$ and $\left[  \mathbf{T}_{\mathfrak{u}}\mathbb{P}_{L}\right]  ^{v}=%
\mathcal{G}%
\oplus%
\mathcal{G}%
_{\perp}$, where 
\begin{eqnarray}%
\mathcal{C}%
_{\perp}:=\bigg\{  \mathbf{X}\in\left[  \mathbf{T}_{\mathfrak{u}}%
  \mathbb{P}_{L}\right]  ^{q}\ &\backslash&\
\left\langle \left.  \mathbf{\Theta}_{q}^{(n)}\right\vert \mathbf{X}%
\right\rangle =0,\>\>  n=1,\dots,N_{0}  \ \bigg\}  ,
\hbox{and}%
\nonumber
\\
\mathcal{G}_{\perp}:=\bigg\{  \mathbf{X}\in\left[
  \mathbf{T}_{\mathfrak{u}}
  \mathbb{P}_{L}\right]  ^{v}\ &\backslash&\
\left\langle \left.  \mathbf{\Theta}_{v}^{(n)}\right\vert \mathbf{X}%
\right\rangle =0, \>\>   n=1,\dots,N_{0}  \ \bigg\}.
\label{TVsplit}%
\end{eqnarray}

To determine the vectors in ker $\mathbf{\Omega}_L(\mathfrak{u})$,
choose a $\mathbf{K}\in\ker \mathbf{\Omega}_{L}(\mathfrak{u})$. Then  
\begin{equation}
\Omega_{M}^{v\flat}\mathbf{K}^{q}=0, \quad \hbox{and}\quad
\Omega_{M}^{q\flat}\mathbf{K}^{v}=\mathbf{-}\Omega_{F}^{\flat}\mathbf{K}^{q}. 
\label{WKV99}%
\end{equation}
Solutions of these equations are found with the use of the following
theorem from linear algebra stated without proof (see also
\cite{Car1988a} where a special case of this theorem was proved).

\begin{theorem}
\label{@Inhomeq}For linear spaces $\mathbf{E}$ and $\mathbf{F}$ of dimension
$D$ and a linear map $\mathcal{A}:\mathbf{F}\rightarrow\mathbf{E}^{\ast}$ of
rank $r$, the inhomogeneous linear equation,%
\begin{equation}
\mathcal{A}\mathbf{f}=\bm{\varphi}, \label{inhomGen}%
\end{equation}
has solutions if and only if%
\begin{equation}
\left\langle \left.  \bm{\varphi}\right\vert \mathbf{e}\right\rangle
=0\ \forall\>\ \mathbf{e}\in\mathbf{A}, \label{cnd11}%
\end{equation}
where%
\begin{equation}
\mathbf{A}:=\left\{  \mathbf{e}\in\mathbf{E}\ \backslash\ \left\langle \left.
\mathcal{A}\mathbf{f}\right\vert \mathbf{e}\right\rangle =0\ \forall\>
\ \mathbf{f}\in\mathbf{F}\ \right\}.  \label{defA}%
\end{equation}
\end{theorem}

This theorem is applied to Eq.~(\ref{WKV99}) by setting $\mathbf{F}=\left[
\mathbf{T}_{\mathfrak{u}}\mathbb{P}_{L}\right]  ^{v}$, $\mathbf{E}=\left[
\mathbf{T}_{\mathfrak{u}}\mathbb{P}_{L}\right]  ^{q}$, $\mathcal{A}%
=\Omega_{M}^{q\flat}$, and $\bm{\varphi}=\mathbf{-}\Omega_{F}^{\flat
}\mathbf{K}^{q}$. To make the connection with the results of
\textbf{Section \ref{&L-Sym}} clear, this application is done
locally. The condition that $\mathbf{X}^{q}\in\mathbf{A}$ is 
$\left\langle \left.  \Omega_{M}^{q\flat}\mathbf{K}\right\vert
\mathbf{X}^{q}\right\rangle =-K^{qa}M_{ab}X^{qb}=0$\ $\forall$ $K^{qa}$, which
requires $M_{ab}X^{qb}=0$. This establishes $\mathbf{A}=%
\mathcal{C}%
$.
Equation (\ref{cnd11}) reduces to $\left\langle \left.  \Omega_{F}^{\flat
}\mathbf{K}^{q}\right\vert \mathbf{C}\right\rangle =F_{ab}K^{qa}%
C^{b}=0\ \forall\>\mathbf{C}\in%
\mathcal{C}%
$,
or equivalently $\mathfrak{z}_{\left(
n\right)  }^{a}F_{ab}K^{qb}=0$. Using the action of
$\Omega^{v\flat}_M$ and $\Omega^{q\flat}_M$ in Eq.~(\ref{WKV99}), we
find that $M_{ab}K^{qa}=0$, and thus $\mathbf{K}^{q}\in%
\mathcal{C}%
$. The existence condition for solutions of Eq.~(\ref{WKV99}) is then,
\begin{equation}
\mathfrak{z}_{\left(  n\right)  }^{a}F_{ab}K^{qb}=0,\ n=1,\ldots
,N_{0}  .\label{NS555}%
\end{equation}
With the basis given in Eq.~$(\ref{basesM})$, we may express
\begin{equation}
  K^{qa} = \sum_{m=1}^{N_0} K^{q(m)} \mathfrak{z}^a_{(m)},
\end{equation}
and Eq.~$(\ref{NS555})$ becomes,
\begin{equation}
\sum_{m=1}^{N_{0}}\bar{F}_{nm}K^{q\left(  m\right)  }=0,\ \text{where }\bar
{F}_{nm}:=\mathfrak{z}_{\left(  n\right)  }^{a}F_{ab}\mathfrak{z}_{\left(
m\right)  }^{b},\label{FbarCbar=0}%
\end{equation}
is the \textbf{reduced matrix} of $F_{ab}$. Then $\mathbf{K}^{q}$
is restricted to the subspace, 
\begin{equation}
\overline{%
\mathcal{C}%
}:=\left\{  \overline{\mathbf{C}}\in%
\mathcal{C}%
\ \backslash\ \sum_{m=1}^{N_{0}}\bar{F}_{nm}\overline{C}^{\left(  m\right)
}=0\right\}  \subset%
\mathcal{C}.%
\label{FCbar}%
\end{equation}

\begin{theorem}
\label{@NVGen}The vectors $\mathbf{K=K}^{q}+\mathbf{K}^{v}\in\ker
\mathbf{\Omega}_{L}$ are given by,%
\begin{equation}
\mathbf{K}^{q}=\overline{\mathbf{C}}\mathbf{,\ \ K}^{v}=\mathbf{G+}%
\widehat{\mathbf{C}},
\label{Kgen}%
\end{equation}
where, $\overline{\mathbf{C}}\in\overline{%
\mathcal{C}%
}$, $\mathbf{G}\in%
\mathcal{G}%
$, and $\widehat{\mathbf{C}}\in%
\mathcal{G}%
_{\perp}$ is the unique solution of $M_{ab}\widehat{C}^{b}=-F_{ab}\overline
{C}^{b}$.

\begin{proof}
The horizontal component, $\mathbf{K}^{q}=\overline{\mathbf{C}}$, of
$\mathbf{K}$ satisfies Eq.~(\ref{NS555}), and is a
solution of Eq.~(\ref{WKV99}). As
$\mathbf{G}\in\ker  \Omega_{M}^{q\flat}  $, the 
general solution of Eq.~(\ref{WKV99}) is $\mathbf{K}^{v}=\mathbf{G+}%
\widehat{\mathbf{C}}$, where 
$\widehat{\mathbf{C}}\in%
\mathcal{G}%
_{\perp}$. With $M_{(\alpha)(\beta)}:=\mathfrak{z}^a_{(\alpha)}M_{ab}\mathfrak{z}^a_{(\beta)}$
for $\alpha,\beta = 1, \dots, D$, and with the choice that
$\mathfrak{z}_{(\alpha)}\in\hbox{ker }M_{ab}(\mathfrak{u})$ for
$\alpha = 1, \dots, N_0$, the components of $\widehat{\mathbf{C}}$ and
$M$ satisfy  $\widehat{C}^{\left(  n\right)  }=0$, and $M_{\left(
  n\right)  \left(   e\right)  }=M_{\left(  f\right)  \left(m\right)
}=0$, when $n,m = 1, \dots, N_0$. Thus 
\begin{equation}
M=\left[
\begin{array}
[c]{cc}%
0 & 0\\
0 & \mathcal{M}%
\end{array}
\right]  ,\ \text{ }\mathcal{M}_{\left(  f\right)  \left(  e\right)
}:=M_{\left(  f\right)  \left(  e\right)  },
\end{equation}
for $e,f=N_{0}+1,\ldots,D$, and $\mathcal{M}$ is
nonsingular. In this basis Eq.~(\ref{WKV99}) becomes $M\widehat{C}=R$,
where $R_{\left(  f\right)  }:=-U_{\left(  f\right) 
}^{va}F_{ab}\overline{C}^{b}$, and $R_{\left(  n\right)  }=0$. The
nonvanishing components  of $\widehat{C}$ and $R$ are vectors with $\left(
D-N_{0}\right)  $-components that satisfy%
\begin{equation}
\mathcal{M}\widehat{C}=R.\label{Gprpsol}%
\end{equation}
Thus there is a unique solution for $\widehat{\mathbf{C}}$ that belongs to $%
\mathcal{G}%
_{\perp}.$
\end{proof}
\end{theorem}

The constant rank assumption for $M_{ab}(\mathfrak{u})$ together with
the definition of $\mathcal{G}$ shows that there are $N_{0}$
free choices for $\mathbf{G}$ at each 
$\mathfrak{u}\in\mathbb{P}_{L}$. According to \textbf{Theorem
  \ref{@NVGen}} the $\widehat{\mathbf{C}}$-term is uniquely specified
by the choice of $\overline{\mathbf{C}}\in\overline{%
\mathcal{C}%
}$. In general, $\bar{F}$ is not
determined by $M_{ab}$, and thus the constant rank of $M_{ab}$ does not
guarantee that the rank of $\bar{F}$ is constant. We then find
\begin{equation}
\dim  \left(\ker  \mathbf{\Omega}_{L}\left(  \mathfrak{u}\right)\right)
  =N_{0}+\bar{D}  ,\ \text{where}%
\ \bar{D}  :=\dim \overline{\mathfrak{C}}\le N_0  .
\label{dim}
\end{equation}

The results of \textbf{Theorem \ref{@NVGen}} are more general than we
need. Because its proof is local and algebraic, even though our focus
is on $\bm{\Omega}_L$ with constant rank the theorem would
nevertheless still hold if the rank was not. It would only have to be
applied to each region of $\mathbb{P}_L$ on which the rank
of $\mathbf{\Omega}_L$ is constant, resulting in a $\bar{D}$ that
takes on different values on the Lagrangian phase space.  

\subsection{Projection of\ $\mathcal{K}$ to ker
  $\mathbf{\Omega}_L(\mathfrak{u})$\label{A-Proj}} 

Consider a region $U_{Sol}\in \mathbb{M}^{(2)}$ on which solutions of the
Euler-Lagrange equations of motion exist, and a point $(t, q, \dot{q},
\ddot{q}) \in U_{Sol}$. Then under the isomorphism $(t, q, \dot{q},
\ddot{q}) \to (t, q, v, X^{va}_{L})$,
$\mathcal{K}\to\mathcal{K}'\subset 
\hbox{\textbf{T}}_{\mathfrak{u}}\mathbb{P}_L$, with a
$\mathbf{k}\in \mathcal{K}$ mapped into a
$\mathbf{k}'\in\mathcal{K}'$ where
\begin{equation}
  \mathbf{k}'=\rho\cdot\frac{\bm{\partial} \>\>\>}{\bm{\partial} q} +
  \dot{\rho}\cdot\frac{\bm{\partial} \>\>\>}{\bm{\partial} v}.
\end{equation}
Now $\dot{\rho} = \mathfrak{L}_{\mathbf{X}_{L}}\rho$, and
$\mathfrak{L}$ is the Lie derivative. From
\textbf{Lemma} $\mathbf{\ref{GS}}$, $\mathcal{K}'\subseteq \hbox{ker
}\mathbf{\Omega}_L(\mathfrak{u})$. But since 
$\dim\> \mathcal{K}=2N_0$, while $\dim \hbox{ ker
  }\mathbf{\Omega}_L(\mathfrak{u})=N_0+\bar{D}\le 2N_0$, it
follows that $\mathcal{K}'=\hbox{ker
}\mathbf{\Omega}_L(\mathfrak{u})$. Although
this conclusion can only be reached on $U_{Sol}$, the rank
of $\mathbf{\Omega}_L$ is constant on $\mathbb{P}_L$, and thus 
$\bar{D}=N_0$ on all of $\mathbb{P}_L$. As such
$\overline{\mathfrak{C}} = \mathfrak{C}$, and for any 
$\mathbf{K}\in \hbox{ker }\mathbf{\Omega}_L(\mathfrak{u})$, $\mathbf{K} =
\mathbf{C} + \mathbf{\dot{C}} + \mathbf{G}$ where now $\dot{\mathbf{C}}
=\dot{C^a}\bm{\partial}/\bm{\partial}v^a$. (This expression for
$\mathbf{K}$ can also be established directly using
Eq.~$(\ref{WKV99})$.) This result is also new.

\subsection{First-order Lagrangian constraints\label{LCon}}

For singular Lagrangians solutions of the energy equation
$\mathbf{X}_E$ are not unique. They also do not, in general, exist
throughout $\mathbb{P}_L$, but are rather confined to a submanifold
of the space given by Lagrangian constraints. The first-order
constraints, those that come directly from the energy equation, are the
focus of this section. Although most of this analysis is done for a
SOLVF, we show later that our results do not depend on
this choice.

With $\mathbf{X}_{L}=\mathbf{X}^q_{L}+\mathbf{X}^v_{L}$, the energy  
equation can be expressed in terms of a one-form $\bm{\Psi}$ as
\begin{equation}
  \Omega_{M}^{q\flat}\mathbf{X}_{L}^{v}=\mathbf{\Psi}.
  \label{EnG0}%
\end{equation}
The existence condition for solutions to Eq.~$(\ref{EnG0})$ is
given again by \textbf{Theorem 
  \ref{@Inhomeq}} by identifying
$\mathcal{A}=\Omega_{M}^{q\flat}$,$\ \mathbf{F}=\left[ 
\mathbf{T}_{\mathfrak{u}}\mathbb{P}_{L}\right]  ^{v}$,\ $\mathbf{E}^{\ast
}=\left[  \mathbf{T}_{\mathfrak{u}}^{\ast}\mathbb{P}_{L}\right]  _{q}$, and
$\mathbf{E}=\left[  \mathbf{T}_{\mathfrak{u}}\mathbb{P}_{L}\right]
^{q}$. Combining Eq.~(\ref{defA}) with the 
action of $\Omega^{q\flat}_M$ on vector fields yields 
$\mathbf{A}=%
\mathcal{C}%
$; consequently Eq.~(\ref{cnd11}) requires that
$\left\langle \left. \mathbf{\Psi}\right\vert \mathbf{C}\right\rangle
=0\ \forall\>\ \mathbf{C}\in \mathcal{C}$ or, after using the
basis $\left\{  \mathbf{U}_{\left(  n\right)  }^{q}\right\} 
$ of $\mathbf{C}$, that $\gamma_{n}^{\left[  1\right]  }:=\left\langle
\left.  \mathbf{\Psi}\right\vert \mathbf{U}_{\left(  n\right)
}^{q}\right\rangle=0$ for $n=1,\ldots,N_{0}$. In terms of local coordinates,
\begin{equation}
\gamma_{n}^{\left[  1\right]  }= U_{\left(  n\right)  }%
^{qa}\left(  \frac{\partial E}{\partial q^{a}}+F_{ab}v^{b}\right).
\end{equation}
The condition $\gamma_{n}^{\left[  1\right]  }=0$ imposes relations on the
coordinates $q$ and $v$, and defines a set of submanifolds of
$\mathbb{P}_L$. These $\gamma_{n}^{\left[1\right]}$ are
called the \textbf{first-order constraint functions}. (Because they are
obtained through the energy equation, they are also called dynamical
constraints in the literature \cite{Mun1992, Car1987a}.)

While the number of first-order
constraint functions in $\hbox{C}_{L}^{\left[  1\right]  }:=\left\{
\gamma_{1}^{\left[  1\right]  },\ldots,\gamma_{N_{0}}^{\left[  1\right]
}\right\}  $ is equal to the dimension of $\mathcal{C}$,
these functions need not be mutually independent. Let $I_{\left[1\right]  }$ be the
number of independent functions in $\hbox{C}_{L}^{\left[  1\right]
}$. Then $I_{\left[ 1\right]  }=\hbox{rank }  \left\{\mathbf{d}\gamma_{n}^{\left[
      1\right]  }\right\} \leq N_{0}$, and 
$\mathbb{P}_{L}^{\left[  1\right]  }:=\left\{ 
\mathfrak{u}\in\mathbb{P}_{L}\ \backslash\ \ \gamma_{n}^{\left[  1\right]
}\left(  \mathfrak{u}\right)  =0\, ,  n=1,\ldots,N_{0}\ \right\}  $\textemdash
called the \textbf{first-order Lagrangian constraint
  submanifold}\textemdash has
$\dim \mathbb{P}_{L}^{\left[1\right]  } =2D-I_{\left[  1\right]
}$. We will assume that $\mathbb{P}_{L}^{\left[ 1\right]  }$ is not
empty, i.e. that the first-order constraint functions are
consistent. Otherwise, there is no SOLVF, and the integral flows that give
the evolution of the dynamical system would not exist.   

While the energy equation is usually written as
$\mathbf{d}E = i_{\mathbf{X}_E}\mathbf{\Omega}_L$, in doing
so we have implicitly restricted ourselves to 
$\mathbb{P}_L^{[1]}$. This is too restrictive for our
purposes, and in this paper we introduce the \textbf{constraint
  one-form}  
\begin{equation}
  \bm{\beta}[\mathbf{X}_E] :=
  \mathbf{d}E-i_{\mathbf{X}_E}\mathbf{\Omega}_L,
  \label{betaform}
\end{equation}
(see also the approach in \cite{Car1987a}). The condition
$\bm{\beta}[\mathbf{X}_E]=0$ then gives both  
solutions of the energy equation and the submanifold
$\mathbb{P}_L^{[1]}$. Furthermore, as $i_{\mathbf{U}^q_{(n)}}\bm{\beta} =
\gamma_n^{[1]}$,   
\begin{equation}
  \left\langle\left(\bm{\beta}[\mathbf{X}_E]- \sum_{m=1}^{N_0}
  \gamma_m^{[1]}\mathbf{\Theta}^{(m)}_q\right)\Bigg\vert
\mathbf{U}^q_{(n)}\right\rangle=0, \quad n=1, \dots, N_0,
\end{equation}
and
\begin{equation}
  \bm{\beta}[\mathbf{X}_E]=\sum_{n=1}^{N_0}
  \gamma_n^{[1]}\mathbf{\Theta}^{(n)}_q +\bm{\vartheta},
\end{equation}
where $\bm{\vartheta}\in\mathbf{T}^{*}_{\mathfrak{u}}\mathbb{P}_L$
such that $\langle\bm{\vartheta}\vert\mathbf{C}\rangle=0$ for all
$\mathbf{C}\in\mathcal{C}$. 
But as $\bm{\beta}[\mathbf{X}_E]=0$ on $\mathbb{P}^{[1]}_L$, we
may choose $\bm{\vartheta}=0$.

From Eq.~$(\ref{betaform})$,
$\bm{\beta}[\mathbf{X}_E+\mathbf{K}]=\bm{\beta}[\mathbf{X}_E]$, and
the construction of $\gamma_n^{[1]}$ does not depend on 
our use of $\mathbf{X}_{L}$.    

\subsection{The Generalized Lie Symmetry Group\label{Sym}} 

Our construction of the generalized Lie
symmetry group for $\mathcal{O}(\mathfrak{u}_0)$ is guided by the
three conditions listed in \textbf{Section 
  \ref{&L-Sym}}, and makes use of the projection of $\mathcal{K}$
to ker $\mathbf{\Omega}_L(\mathfrak{u})$ in \textbf{Section
  \ref{A-Proj}}. It begins with the vector space,
\begin{equation}
  \overline{\hbox{ker }\mathbf{\Omega}_L(\mathfrak{u})} :=
  \{\mathbf{P} \in \hbox{ker } \mathbf{\Omega}_L(\mathfrak{u})\
  \backslash \ [\mathbf{G},\mathbf{P}]
  \in\left[\mathbf{T}_{\mathfrak{u}}\mathbb{P}_L\right]^v \>\>\forall
  \>\>\mathbf{G}\in\mathcal{G}\},  
\end{equation}
along with the following collection of functions on $\mathbb{P}_L$,  
\begin{equation}
  \overline{\mathcal{F}} := \{f\in C^\infty \hbox{on } \mathbb{P}_L
  \ \backslash\ \ \mathbf{G}f = 0 \>\>\forall\>\> \mathbf{G} \in
  \mathcal{G}\}.
  \label{function}
\end{equation}
The following result will be used a number of times in our analysis. 
\begin{lemma} \label{basic} Let  
  $\mathbf{X}\in\mathbf{T}_{\mathfrak{u}}\mathbb{P}_L$ 
  and $\mathbf{G}\in\mathcal{G}$ such that $[\mathbf{G},
    \mathbf{X}]\in\hbox{ker }\bm{\Omega}_L(\mathfrak{u})$. Then 
  $[\mathbf{G},
    \mathbf{X}]\in\left[\mathbf{T}_{\mathfrak{u}}\mathbb{P}_L\right]^v$
  iff $[\mathbf{G}, \mathbf{X}]\in\mathcal{G}$.

  \begin{proof} Since $[\mathbf{G}, \mathbf{X}]\in\hbox{ker
    }\bm{\Omega}_L(\mathfrak{u})$, from \textbf{Theorem
      \ref{@NVGen}}  there exists a $\mathbf{C}\in\mathfrak{C}$ and
    $\mathbf{G}'\in\mathcal{G}$ such that $[\mathbf{G},
      \mathbf{X}]=\mathbf{C} + \widehat{\mathbf{C}} +\mathbf{G}'$, and
    we see that $[\mathbf{G}, \mathbf{X}]\in
    \left[\mathbf{T}_{\mathfrak{u}}\mathbb{P}_L\right]^v$ iff 
    $\mathbf{C}=0$. Then $\widehat{\mathbf{C}}=0$, 
    and $[\mathbf{G}, \mathbf{X}]\in\mathcal{G}$.
  \end{proof}
\end{lemma}
It then follows that $[\mathbf{G},\mathbf{P}]\in\mathcal{G}$ for all
$\mathbf{P}\in \overline{\hbox{ker
  }\bm{\Omega}_L(\mathfrak{u})}$. As $\mathcal{G}$ is involutive
and as $\mathcal{G}\subset \hbox{ker }\bm{\Omega}_L(\mathfrak{u})$,
$\mathcal{G}\subset \overline{\hbox{ker
  }\bm{\Omega}_L(\mathfrak{u})}$ as well, and thus  
$\mathcal{G}$ is an ideal of $\overline{\hbox{ker
  }\bm{\Omega}_L(\mathfrak{u})}$.  

\begin{lemma}
  \label{basis}
  There exists a choice of basis for ker
  $\mathbf{\Omega}_L(\mathfrak{u})$ that is 
  also a basis of $\overline{\hbox{ker
    }\mathbf{\Omega}_L(\mathfrak{u})}$. 

  \begin{proof} 
  Given a basis $\{\mathbf{G}_{(n)}\}$ of $\mathcal{G}$, choose
  a set $\{\mathbf{K}_{(n)}\}$ such that $\{\mathbf{K}_{(n)},
  \mathbf{G}_{(n)}, n=1, \dots, N_0\}$ form a basis of 
  ker $\mathbf{\Omega}_L(\mathfrak{u})$. These
  $\{\mathbf{G}_{(n)}\}$ are also a basis of
  $\overline{\hbox{ker }\mathbf{\Omega}_L(\mathfrak{u})}$. To show that
    $\{\mathbf{K}_{(n)}\}$ can be chosen to complete this basis, express
  $\mathbf{K}_{(n)} = \mathbf{C}_{(n)} + 
  \widehat{\mathbf{C}}_{(n)}+\mathbf{G}_{(n)}$. Then as
  $[\mathbf{G}_{(m)}, \mathbf{K}_{(n)}] = 
    [\mathbf{G}_{(m)}, \mathbf{C}_{(n)}]+ [\mathbf{G}_{(m)},
      \widehat{\mathbf{C}}_{(n)} 
      +\mathbf{G}_{(n)}]$, we need only show that there exists a
    choice of 
  $\{\mathbf{C}_{(n)}\}$ such that $[\mathbf{G}_{(m)},
    \mathbf{C}_{(n)}]\in[\mathbf{T}_{\mathfrak{u}}\mathbb{P}_L]^v$. This
  we do by construction.  
  
  Because ker $\mathbf{\Omega}_L(\mathfrak{u})$ is involutive,
  $[\mathbf{G}_{(m)}, \mathbf{K}_{(n)}]\in\hbox{ker
  }\mathbf{\Omega}_L(\mathfrak{u})$, and there 
  exist functions $\lambda_{mn}^{\quad\> l}$ on $\mathbf{P}_L$ such that
  \begin{equation}
   \mathbf{T}\tau_{\mathbb{Q}} [\mathbf{G}_{(m)},
    \mathbf{C}_{(n)}] = \sum_{l=0}\lambda_{mn}^{\quad\> l}
    \mathbf{C}_{(l)}.
    \label{e12}
  \end{equation}
  Let $\{\underline{\mathbf{C}}_{(n)}, n=1, \dots, N_0\}$ be another
  choice of basis of $\mathcal{C}$ where
  \begin{equation}
    \underline{\mathbf{C}}_{(n)} = \sum_{m=1}^{N_0}
    \omega_n^{\>m}\mathbf{C}_{(m)}.
  \end{equation}
  Requiring 
  $\mathbf{T}\tau_{\mathbb{Q}} [\mathbf{G}_{(m)},
    \underline{\mathbf{C}}_{(n)}]=0$ in turn requires that
  $\omega_n^{\>m}$ be a solution of 
  \begin{equation}
    \mathbf{G}_{(l)}\omega_n^{\>m}  + \sum_{k=1}^{N_0}
    \omega_n^{\>k}\lambda_{lk}^{\>\>\> m}=0.
    \label{omega}
  \end{equation}
  This is a linear, first-order Cauchy problem \cite{Lee2013}. A
  solution exists for a given set of boundary conditions for
  $\omega_n^{\>m}$ given on a surface $\mathcal{S}$ as long as 
  $\mathbf{G}_{(l)}$ is nowhere tangent to $\mathcal{S}$ 
  \cite{Lee2013}. As $N_0<D$, and as we have complete freedom to
  choose both the boundary conditions and $\mathcal{S}$, a
  solution of Eq.~$(\ref{omega})$ can always be found. The collection
  of vector fields $\{\mathbf{P}_{(n)} = \underline{\mathbf{C}}_{(n)}
  + \widehat{\underline{\mathbf{C}}}_{(n)}, \mathbf{G}_{(n)}\}$ is
  then a basis of both ker $\mathbf{\Omega}_L(\mathfrak{u})$ and
  $\overline{\hbox{ker }\mathbf{\Omega}_L(\mathfrak{u})}$.   
  \end{proof}
\end{lemma}
For the rest of this paper we will assume that this choice of
basis for $\mathcal{C}$ and $\overline{\hbox{ker
  }\mathbf{\Omega}_L(\mathfrak{u})}$ has been made.  

It follows from \textbf{Lemma} \ref{basis} that dim
$\left(\overline{\hbox{ker }\mathbf{\Omega}_L(\mathfrak{u})}\right) =
2N_0$. Next, choose two vectors $\mathbf{P}_{1, 2}\in
  \overline{\hbox{ker }\mathbf{\Omega}_L(\mathfrak{u})}$. Then
  $\mathbf{P}_{1,2}\in\hbox{ker }\bm{\Omega}_L(\mathfrak{u})$ as well,
  and as ker $\mathbf{\Omega}_L(\mathfrak{u})$ is involutive,
  $[\mathbf{P}_1, \mathbf{P}_2]\in \hbox{ker
  }\mathbf{\Omega}_L(\mathfrak{u})$. Choose now a
  $\mathbf{G}\in\mathcal{G}$. From the Jacobi identity,
  $[\mathbf{G}, [\mathbf{P}_1, \mathbf{P}_2]]=- [\mathbf{P}_1,
    [\mathbf{P}_2, \mathbf{G}]]-[\mathbf{P}_2, [\mathbf{G},
      \mathbf{P}_1]]$. From \textbf{Lemma \ref{basic}}, 
  there exists $\mathbf{G}_{1,2}\in\mathcal{G}$ such that
  $\mathbf{G}_{1,2}=[\mathbf{G}, \mathbf{P}_{1,2}]$. Then
  $[\mathbf{G}, [\mathbf{P}_1, \mathbf{P}_2]]= [\mathbf{P}_1,
    \mathbf{G}_2]-[\mathbf{P}_2, \mathbf{G}_1]$, and thus
  $\overline{\hbox{ker }\mathbf{\Omega}_L(\mathfrak{u})}$ is 
  involutive. 

As $\mathcal{G}$ is an ideal of $\overline{\hbox{ker
  }\mathbf{\Omega}_L(\mathfrak{u})}$, we may define for
any $\mathbf{P}_1, \mathbf{P}_2\in\overline{\hbox{ker 
  }\mathbf{\Omega}_L(\mathfrak{u})}$ the equivalence relation:
$\mathbf{P}_1\sim\mathbf{P}_2$ iff
$\mathbf{P}_1-\mathbf{P}_2\in\mathcal{G}$. The equivalence class,
\begin{equation}
  \left[\mathbf{P}\right]:=
  \{\mathbf{Y}\in \overline{\hbox{ker
    }\mathbf{\Omega}_L(\mathfrak{u})}\ \backslash 
  \ \mathbf{Y}\sim\mathbf{P}\},  
\end{equation}
then follows, along with the quotient space
$\overline{\hbox{ker }\mathbf{\Omega}_L(\mathfrak{u})}/\mathcal{G}$.   
This results in a collection of vectors that lie in the kernel of
$\mathbf{\Omega}_L$, but with the vectors in $\mathcal{G}$
removed. This 
$\overline{\hbox{ker 
  }\mathbf{\Omega}_L(\mathfrak{u})}/\mathcal{G}$
addresses the first two conditions listed at the end of \textbf{Section
  \ref{&L-Sym}}. We now turn our attention to the third condition.  

Because the integral flow $\mathfrak{u}_{\mathbf{X}}(t)$ of any solution
$\mathbf{X}$ of the energy equation must lie on $\mathbb{P}_L^{[1]}$, a 
symmetry transformation of $\mathfrak{u}_{\mathbf{X}}(t)$ must result
in an integral flow $\mathfrak{u}_{\mathbf{Y}}(t)$ of another solution
$\mathbf{Y}$ of the energy equation. This flow must also lie on
$\mathbb{P}_L^{[1]}$. Implementing this condition is done through
$\bm{\beta}[\mathbf{X}_E]$, and it is this 
one-form that singles out the vectors in ker 
$\mathbf{\Omega}_L(\mathfrak{u})$ that generate the
generalized Lie symmetry. We do this by looking at the action
of a vector $\mathbf{G}\in\mathcal{G}$ on
$\bm{\beta}[\mathbf{X}_E]$. 

As $i_{\mathbf{G}}\bm{\beta}[\mathbf{X}_E]=0$ for all
$\mathbf{G}\in\mathcal{G}$, 
\begin{equation}
  \mathfrak{L}_{\mathbf{G}}\bm{\beta}[\mathbf{X}_E]= \sum_{n=1}^{N_0}
  \left(\mathbf{G}\gamma_n^{[1]}\right) \mathbf{\Theta}^{(n)}_q,
\end{equation}
since $\gamma_n^{[1]}=0$ on $\mathbb{P}_L^{[1]}$. Given a
$\mathbf{P}_{(n)} \in\overline{\hbox{ker
  }\bm{\Omega}_L(\mathfrak{u})}$ such that $\mathbf{P}_{(n)} = 
\mathbf{U}_{(n)}^q + \widehat{\mathbf{U}}_{(n)} +\mathbf{G}'$ with
$\mathbf{G}'\in\mathcal{G}$, it has been shown in \cite{Got1979,
  Got1980} that $\gamma_n^{[1]}=i_{\mathbf{P}_{(n)}}\mathbf{d}E$. Then
  $\mathbf{G}\gamma_n^{[1]}= [\mathbf{G}, \mathbf{P}_{(n)}]E +
\mathbf{P}_{(n)}\mathbf{G}E$. But $\mathcal{G}$ is an ideal of
$\overline{\hbox{ker }\bm{\Omega}_L(\mathfrak{u})}$, while a
straightforward calculation shows that $\mathbf{G}E=0$. Thus
$\mathbf{G}\gamma_{(n)}^{[1]}=0$, and it follows that 
$\mathfrak{L}_{\mathbf{G}}\bm{\beta} =0$. The subspace   
\begin{equation}
  \mathcal{S}\hbox{ym} := \big\{\mathbf{P}\in \overline{\hbox{ker
    }\mathbf{\Omega}_L(\mathfrak{u})}/\mathcal{G}
  \ \backslash\ \
  \mathfrak{L}_{\mathbf{P}}\bm{\beta}[\mathbf{X}_E] =
  \mathbf{d}(i_{\mathbf{P}}\bm{\beta}[\mathbf{X}_E]) \hbox{ on }
  \mathbb{P}_L^{[1]}\big\}, 
\end{equation}
is then well defined, and we find that
$\mathbf{P}\in\mathcal{S}\hbox{ym}$ iff   
$i_{\mathbf{P}}\mathbf{d}\bm{\beta}[\mathbf{X}_E]=0$. A
simple calculation shows 
that $\mathcal{S}\hbox{ym}$ is involutive. There is then
a corresponding set of one-parameter subgroups 
$\mathbf{\sigma}_{\mathbf{P}}(\epsilon,x)$ for any $\mathbf{P}\in 
\mathcal{S}\hbox{ym}$ given by
\begin{equation}
  \frac{d\mathbf{\sigma}_{\mathbf{P}}}{d\epsilon} :=
  \mathbf{P}\left(\mathbf{\sigma}_{\mathbf{P}}\right), 
\end{equation}
with $\sigma_{\mathbf{P}}(0,\mathfrak{u}) = \mathfrak{u}$ for
$\mathfrak{u}\in\mathbb{P}_L$. The collection of such 
subgroups gives the Lie group
$\hbox{Gr}_{\mathcal{S}\hbox{ym}}$. This
$\hbox{Gr}_{\mathcal{S}\hbox{ym}}$ is the generalized  
symmetry group we are looking for, as we see below.

\subsection{Euler-Lagrange Solutions of the Energy
  Equation\label{Sol}}

The set of \textbf{general solutions} to the energy equation is 
\begin{equation}
\mathcal{S}\hbox{ol}
:=\{\mathbf{X}_{E}\in\mathbf{T}_{\mathfrak{u}}\mathbb{P}_L
\ \backslash\ \ i_{\mathbf{X}_{E}}\mathbf{\Omega}_L = \mathbf{d}E
\hbox{ on } \mathbb{P}_L^{[1]}\}.
\label{GenSol}
\end{equation}
Importantly, while a SOLVF
$\mathbf{X}_{L}\in\mathcal{S}\hbox{ol}$, the 
majority of vectors in $\mathcal{S}\hbox{ol}$ are \textit{not}
SOLVFs. This is the root cause of the ``second  
order problem'' first raised by K\"unzle \cite{Kun1969} (see
also \cite{Got1979}, \cite{Got1980}, and \cite{Car1988a}).

If $\mathfrak{u}(t)$ is the integral flow of a vector in
$\mathcal{S}\hbox{ol}$ whose projection onto
$\mathbb{Q}$ corresponds to a trajectory $q(t)$ that is a solution of
the Euler-Lagrange equations of motion, then
$\hbox{Gr}_{\mathcal{S}\hbox{ym}}$ must map one of
such flows into another. However, while
$\mathfrak{L}_{\mathbf{G}}\mathbf{X}_{L} = 
[\mathbf{G}, \mathbf{X}_{L}]\in \hbox{ker
}\mathbf{\Omega}_L(\mathfrak{u})$, in general
$\mathfrak{L}_{\mathbf{G}}\mathbf{X}_{L} \notin \mathcal{G}$. The 
action of $\sigma_{\mathbf{P}}$ on the flow
$\mathfrak{u}_{\mathbf{X}_{L}}$ will in general result in a flow
$\mathfrak{u}_{\mathbf{Y}}$ generated by a $\mathbf{Y}$ that is
\textit{not} a SOLVF. It need not even be a solution of the energy
equation. By necessity, general solutions of the energy equation must
be considered. Only a specific subset of such solutions are 
physically relevant, however.    

As 
\begin{equation}
  i_{[\mathbf{X}_E,
      \mathbf{P}]}\mathbf{\Omega}_L=i_{\mathbf{P}}\mathbf{d}\bm{\beta}[\mathbf{X}_E],
  \label{Cond}
\end{equation}
for $\mathbf{P}\in\hbox{ker }\mathbf{\Omega}_L(\mathfrak{u})$ and 
$\mathbf{X}_E\in\mathcal{S}\hbox{ol}$, in general
$\mathfrak{L}_{\mathbf{P}}\mathbf{X}_E 
\notin\hbox{ker }\bm{\Omega}_L(\mathfrak{u})$. The exception is when
$\mathbf{P}\in\mathcal{S}\hbox{ym}$ as well, which leads
us to the subset of solutions 
\begin{equation}
\overline{\mathcal{S}\hbox{ol}}
:=\{\overline{\mathbf{X}}_{EL}\in\mathcal{S}\hbox{ol}  
\ \backslash \ [\mathbf{G},
  \overline{\mathbf{X}}_{EL}]\in
\left[\mathbf{T}_{\mathfrak{u}}\mathbb{P}_L\right]^v \>\>
\>\> \forall \mathbf{G}\in
\mathcal{G}\}.  
\end{equation}
Moreover, as $i_{[\mathbf{G},\overline{\mathbf{X}}_{EL}]}\mathbf{\Omega}_L =
-\mathfrak{L}_{\mathbf{G}}\bm{\beta}=0$, from \textbf{Lemma
  \ref{basic}}
$[\mathbf{G},\overline{\mathbf{X}}_{EL}]\in\mathcal{G}$.  

\begin{lemma} \label{X_EL} $[\overline{\mathbf{X}}_{EL}, \mathbf{P}] \in
  \overline{\hbox{ker } \mathbf{\Omega}_L(\mathfrak{u})}$ for all
  $\mathbf{P}\in \mathcal{S}\hbox{ym}$.  

  \begin{proof} As $\mathbf{P}\in\mathcal{S}\hbox{ym}$, 
    from Eq.~$(\ref{Cond})$ $[\overline{\mathbf{X}}_{EL},
      \mathbf{P}]\in\hbox{ker }
    \mathbf{\Omega}_L(\mathfrak{u})$. Next, for each
    $\mathbf{G}\in\mathcal{G}$, there is a
    $\mathbf{G}_{\mathbf{X}}\in\mathcal{G}$ such that 
    $\mathbf{G}_{\mathbf{X}}=[\mathbf{G},\overline{\mathbf{X}}_{EL}]$.  
    There is also a $\mathbf{G}_{\mathbf{P}}\in\mathcal{G}$ such that 
    $\mathbf{G}_{\mathbf{P}}=[\mathbf{G},\mathbf{P}]$. It then follows
    from the Jacobi identity that
      $[[\overline{\mathbf{X}}_{EL}, 
      \mathbf{P}],\mathbf{G}]\in\mathcal{G}$, and 
    $[\overline{\mathbf{X}}_{EL}, \mathbf{P}] \in \overline{\hbox{ker
      }\mathbf{\Omega}_L(\mathfrak{u})}$. 
\end{proof}
\end{lemma}
The vector fields in $\overline{\mathcal{S}\hbox{ol}}$
generate the family of integral flows  
\begin{equation}
  \mathcal{O}_{EL}(\mathfrak{u}_0) := \bigg\{\mathfrak{u}(t) \ \backslash \
  \frac{d\mathfrak{u}}{dt}=\overline{\mathbf{X}}_{EL}(\mathfrak{u}),
  \overline{\mathbf{X}}_{EL}\in\overline{\mathcal{S}\hbox{ol}},
  \>\>\hbox{and }
  \mathfrak{u}(t_0)=\mathfrak{u}_0\bigg\}.
\end{equation}
The physical significance of these flows can be seen from the
following theorem. 

\begin{theorem} $\hbox{Gr}_{\mathcal{S}\hbox{ym}}$ forms a group of
  symmetry transformations of $\mathcal{O}_{EL}(\mathfrak{u}_0)$.
  
  \begin{proof}
    Let $\mathfrak{u}_{\overline{\mathbf{X}}_{EL}}(t,\mathfrak{u}_0)\in
    \mathcal{O}_{EL}(\mathfrak{u}_0)$ be an integral flow generated by
    $\overline{\mathbf{X}}_{EL}$, and let
    $\mathbf{\sigma}_{\mathbf{P}}(\epsilon,\mathfrak{u})\in\hbox{Gr}_{\mathcal{S}\hbox{ym}}$
    be a one-parameter subgroup of $\hbox{Gr}_{\mathcal{S}\hbox{ym}}$
    generated by $\mathbf{P}\in\mathcal{S}\hbox{ym}$ with
    $\sigma_{\mathbf{P}}(0, \mathfrak{u})=\mathfrak{u}$. The action of 
    $\mathbf{\sigma}_{\mathbf{P}}$ on
    $\mathfrak{u}_{\overline{\mathbf{X}}_E}$ gives
    $\mathfrak{u}_{\mathbf{Y}}(t,\mathfrak{u}_0):=\mathbf{\sigma}_{\mathbf{P}}(\epsilon,  
      \mathfrak{u}_{\overline{\mathbf{X}}_{EL}}(t,\mathfrak{u}_0))$,
      while the choice $\epsilon=0$ when $t=t_0$ ensures that 
    $\mathfrak{u}_{\overline{\mathbf{X}}_{EL}}$ and
      $\mathbf{u}_{\mathbf{Y}}$ have the same initial data. The tangent
    to this path is 
    \begin{equation}
      \mathbf{Y}=\mathbf{\sigma}_{\mathbf{P}}^{*}\circ
      \overline{\mathbf{X}}_{EL}(\sigma^{-1}_{\mathbf{P}}(\epsilon,
      \mathfrak{u}_{\overline{\mathbf{X}}_{EL}})), 
    \end{equation}
    where $\mathbf{\sigma}_{\mathbf{P}}^{*}$ is the pullback map of
    $\mathbf{\sigma}_{\mathbf{P}}$. As 
    $\mathbf{\sigma}_{\mathbf{P}}$ is also a mapping of 
    $\mathbb{P}_L$ into itself, for a suitably small neighborhood about
    $\mathfrak{u}_{\overline{\mathbf{X}}_{EL}}$ we may expand $\mathbf{Y}$ about
    $\epsilon=0$ in the Lie series, 
    \begin{equation}
      \mathbf{Y}(\mathbf{\sigma}(\epsilon,\mathfrak{u}_{\overline{\mathbf{X}}_{EL}}(t, 
      \mathfrak{u}_0))) =\sum_{n=0}^\infty \frac{\epsilon^n}{n!}
      \mathfrak{L}_{\mathbf{P}}^{(n)}
      \overline{\mathbf{X}}_E\Big\vert_{\mathfrak{u}_{\overline{\mathbf{X}}_{EL}}(t,\mathfrak{u}_0)}. 
    \end{equation}
    However, from \textbf{Lemma $\ref{X_EL}$},
      $\mathfrak{L}_{\mathbf{P}}\overline{\mathbf{X}}_{EL} \in 
    \overline{\hbox{ker }\mathbf{\Omega}_L(\mathfrak{u})}$, and
    $\overline{\hbox{ker }\mathbf{\Omega}_L(\mathfrak{u})}$ is
    involutive. Then   
    $\mathbf{Y}(\sigma_{\mathbf{P}}(\epsilon,\mathfrak{u}_{\overline{\mathbf{X}}_{EL}}(t,
    \mathfrak{u}_0))) =
    \overline{\mathbf{X}}_{EL}(\mathfrak{u}_{\overline{\mathbf{X}}_{EL}}(t, 
    \mathfrak{u}_0)) + \epsilon\overline{\mathbf{Z}}(\epsilon,
      \mathfrak{u}_{\overline{\mathbf{X}}_{EL}}(t, 
      \mathfrak{u}_0))$, where $\overline{\mathbf{Z}}\in \overline{\hbox{ker
      }\mathbf{\Omega}_L(\mathfrak{u})}$. It then follows that
      $\mathbf{Y}\in\overline{\mathcal{S}\hbox{ol}}$,
      and $\mathfrak{u}_{\mathbf{Y}}(t, \mathfrak{u}_0)\in
      \mathcal{O}_{EL}(\mathfrak{u}_0)$.  
  \end{proof}
\end{theorem}

While $\mathbf{X}_{L}\notin \overline{\mathcal{S}\hbox{ol}}$, it is
possible to construct from $\mathbf{X}_{L}$ a vector field
$\overline{\mathbf{X}}_{L}$ that is. Choose a basis 
$\left\{\mathbf{P}_{(n)}, \mathbf{G}_{(n)}, n = 1, 
\dots, N_0\right\}$ of $\overline{\hbox{ker
  }\mathbf{\Omega}_L(\mathfrak{u})}$, and consider a vector field 
$\overline{\mathbf{X}}_{L}$ such that 
\begin{equation}
  \overline{\mathbf{X}}_{L} = \mathbf{X}_{L} + \sum_{m=1}^{N_0}
  f^m(\mathfrak{u}) \mathbf{P}_{(m)} +\mathbf{G},
\end{equation}
where $\mathbf{G}\in\mathcal{G}$, and $f^m(\mathfrak{u})$ are functions on
$\mathbb{P}_L$. For
$\overline{\mathbf{X}}_{L}\in\overline{\mathcal{S}\hbox{ol}}$ as well,
we must have $[\overline{\mathbf{X}}_{L}, \mathbf{G}_{(n)}]\in
\mathcal{G}$, and thus these functions must be solutions of 
\begin{equation}
  \mathbf{G}_{(n)}f^m(\mathfrak{u}) =- \left\langle
  \mathbf{\Theta}^{(m)}_q\big\vert 
         [\mathbf{X}_{L}, \mathbf{G}_{(n)}] \right\rangle.
  \label{f}
\end{equation}
Once again, this is a linear Cauchy problem, and a solution exists
with the appropriate choice of boundary conditions and surfaces. 

If $f^m(\mathfrak{u})$ is a solution to Eq.~$(\ref{f})$, then
$f^m(\mathfrak{u})+u^m(\mathfrak{u})$ 
is as well as long as $u^m(\mathfrak{u})\in\overline{\mathcal{F}}$. This
leads us to the \textbf{second-order, Euler-Lagrange vector
  field} (SOELVF),  
\begin{equation}
  \overline{\mathbf{X}}_{EL}=\overline{\mathbf{X}}_{L} +
  \sum_{m=1}^{N_0} u^m(\mathfrak{u}) \left[\mathbf{P}_{(m)}\right],
  \label{EL}
\end{equation}
where $\{[\mathbf{P}_{(n)}], n=1, \dots, N_0\}$ is a choice of basis
for $\overline{\hbox{ker
  }\mathbf{\Omega}_L(\mathfrak{u})}/\mathcal{G}$. By construction,
$\overline{\mathbf{X}}_{EL}\in
\overline{\mathcal{S}\hbox{ol}}$. Conversely, if
$\overline{\mathbf{Y}}_{EL} 
\in \overline{\mathcal{S}\hbox{ol}}$, then 
$\overline{\mathbf{Y}}_{EL}-\overline{\mathbf{X}}_{L} \in
\overline{\hbox{ker }\mathbf{\Omega}_L(\mathfrak{u})}$ and
$\overline{\mathbf{Y}}_{EL}$ is a SOELVF. Thus,
$\overline{\mathbf{Y}}_{EL}\in \overline{\mathcal{S}\hbox{ol}}$ iff 
$\overline{\mathbf{Y}}_{EL}$ is a SOELVF. 

\subsection{A constraint algorithm for second-order, Euler-Lagrange
  vector fields\label{&StabC}}

For most dynamical systems the flow fields in
$\mathcal{O}_{EL}(\mathfrak{u}_0)$ will not 
be confined to $\mathbb{P}_{L}^{\left[  1\right]
}$, and yet this is  the submanifold on which the solutions 
$\overline{\mathbf{X}}_{EL}\in\overline{\mathcal{S}\hbox{ol}}$ of the energy
equations exist. In these cases it is necessary to 
jointly choose a SOELVF $\overline{\mathbf{X}}_{EL}$ and a submanifold
of $\mathbb{P}_{L}^{\left[  1\right]  }$ on which 
$\mathfrak{u}_{\overline{\mathbf{X}}_{EL}}$ will be confined. Doing so requires that  
\begin{equation}
  \mathfrak{L}_{\overline{\mathbf{X}}_E}\bm{\beta}=0,
  \label{stable}
\end{equation}
which is called the \textbf{constraint condition}. Implementing
it involves imposing successive 
conditions on $\overline{\mathbf{X}}_{EL}$. At each step
additional constraints may be introduced, giving a
succession of submanifolds of $\mathbb{P}_{L}^{\left[  1\right]  }$. It is an
iterative process that terminates either when
$\mathfrak{u}_{\overline{\mathbf{X}}_{EL}}$  
is confined to the current submanifold under the current generator of time
evolution, or when the possibility of dynamics on $\mathbb{P}_L$ is
exhausted. This process is called a constraint algorithm, and has been
introduced often in the literature. While such an algorithm will
also be presented here, its purpose is to show that the end result
$\overline{\mathbf{X}}_{EL}$ of the algorithm is once again a SOELVF,
and a second-order problem is avoided. Later, it will also be used
to show that both this $\overline{\mathbf{X}}_{EL}$ and the Lagrangian
constraints\textemdash whether first-order or introduced
by the algorithm\textemdash are projectable.   

To present the constraint algorithm we
introduce the following notation used in conjection with the
constraint analysis
\begin{equation}
\overline{\mathbf{X}}_{EL}^{[1]} := \overline{\mathbf{X}}_{EL}, \>
\overline{\mathbf{X}}_{L}^{[1]} := \overline{\mathbf{X}}_{L}, \>
\mathbf{P}^{[1]}_{(n)} := \mathbf{P}_{(n)}, \> u_{[1]}^m := u^m, \>
N_0^{[1]} := N_0.
\label{notation}
\end{equation}
As both $u_{[1]}^n,
\gamma^{[1]}_n\in \overline{\mathcal{F}}$,
$\left[\mathbf{P}_{(n)}^{[1]}\right]\gamma^{[1]}_m = 
\mathbf{P}_{(n)}\gamma^{[1]}_m$. The constraint condition
Eq.~$(\ref{stable})$ requires 
$\mathfrak{L}_{\overline{\mathbf{X}}_E}\gamma_n^{[1]}=0$, which, after
using Eq.~$(\ref{EL})$ for a general SOELVF, reduces to
\begin{equation}
  \sum_{m=1}^{N_0} \Gamma^{[1]}_{nm} u^m_{[1]} =
  -\left\langle \mathbf{d} \gamma^{[1]}_n\Big\vert
  \overline{\mathbf{X}}^{[1]}_{L}\right\rangle, \>\hbox{with }
  \Gamma^{[1]}_{nm} := \left\langle
  \mathbf{d}\gamma^{[1]}_n\Big\vert \mathbf{P}^{[1]}_{(m)}\right\rangle.
  \label{first-order}
\end{equation}
Then $r^{[1]} :=\hbox{rank } \Gamma^{[1]}_{nm}$ of
the $u^m_{[1]}$ is determined by Eq.$(\ref{first-order})$, while
$N_0^{[2]} :=N_0^{[1]}-r^{[1]}$ are not. Moreover, $N_0^{[2]}$
\textbf{second-order Lagrangian constraint functions} 
\begin{equation}
  \gamma^{[2]}_{n_{[2]}} := \left\langle
  \mathbf{d}\gamma^{[1]}_{n_{[2]}}\Big\vert 
  \overline{\mathbf{X}}_{L}^{[1]}\right\rangle, n_{[2]}=1, \cdots,
  N_0^{[2]}. 
  \label{2nd}
\end{equation}
are introduced with the conditions $\gamma^{[2]}_{n_{[2]}}=0$
imposed. In general there will be $I_{[2]}:=
\hbox{rank }\left\{\mathbf{d}\gamma^{[1]}_{n_{[1]}},
  \mathbf{d}\gamma^{[2]}_{n_{[2]}}\right\}$
independent functions in $\hbox{C}^{[2]}_L :=
\hbox{C}^{[1]}\cup \left\{\gamma^{[2]}_{n_{[2]}}\ \backslash
\ n_{[2]} = 1, \dots, N_0^{[2]}\right\}$, and $\mathbb{P}_L^{[1]}$ is
reduced to the \textbf{second-order constraint submanifold}, 
\begin{equation}
  \mathbb{P}_L^{[2]} := \left\{\mathfrak{u}\in\mathbb{P}_L^{[1]}
  \ \backslash \ \gamma^{[2]}_{[n_2]}=0, n_{[2]} = 1, \dots,
  N_0^{[2]} \right\}, 
\end{equation}
where dim $\mathbb{P}^{[2]}_L = 2D-I_{[2]}$. At this point, there are two
possibilities. If $I_{[2]}=I_{[1]}$ or
$I_{[2]}=2D$, the iterative process stops, and no new Lagrangian
constraints are introduced. If not, the process continues. 

For the second step in the iterative process, we choose a basis
$\left\{[\mathbf{P}_{(n)}^{[2]}]\right\}$ for $\overline{\hbox{ker
  }\mathbf{\Omega}_L(\mathfrak{u})}/\mathcal{G}$ and arbitrary functions
$\left\{u_{[2]}^m\right\}$ such that for $m=1, \dots, N_0^{[2]}$,
$u_{[2]}^m$ are linear combinations of $u_{[1]}^m$ that lie in the kernel
$\Gamma^{[1]}_{nm}$.
Then
\begin{equation}
\overline{\mathbf{X}}_{EL}^{[2]} = \overline{\mathbf{X}}_{L}^{[2]} +
\sum_{m=1}^{N_0^{[2]}} u_{[2]}^m \left[\mathbf{P}_{(m)}^{[2]}\right],
\end{equation}
with
\begin{equation}
  \overline{\mathbf{X}}_{L}^{[2]} = \overline{\mathbf{X}}_{L}^{[1]} +
  \sum_{m=N_0^{[2]}+1} ^{N_0^{[1]}}u^m_{[2]}\left[\mathbf{P}^{[2]}_{(m)}\right].
\end{equation}
Here, the functions $u^m_{[2]}$ for $m = N_0^{[2]}+1, \dots, N_0^{[1]}$ have been
determined through the constraint analysis of $\gamma^{[1]}_n$. 

Because for any $\mathbf{G}\in\mathcal{G}$,
$\mathbf{G}i_{\overline{\mathbf{X}}_E^{[1]}}\mathbf{d}\gamma^{[1]}_n= 
  \mathfrak{L}_{[\mathbf{G},\overline{\mathbf{X}}_E^{[1]}]}\gamma^{[1]}_n
  =0$ and $\mathbf{G}\Gamma^{[1]}_{nm} =
  \mathfrak{L}_{[\mathbf{G},\mathbf{P}_m^{[1]}]}\gamma^{[1]}_n=0$, it
  follows that $\mathbf{G}u_{[1]}^m=0$, as required. Similarly,
  $\mathbf{G}\gamma^{[2]}_n = \mathfrak{L}_{[\mathbf{G},
      \overline{\mathbf{X}}_{EL}]}\mathbf{d}\gamma^{[2]}_n=0$. Clearly
  $\gamma^{[2]}_n\in\overline{\mathcal{F}}$ and we may require
  $u^m_{[2]}\in\overline{\mathcal{F}}$ as well. It 
  then follows that $\left[\mathbf{P}^{[2]}_{(n)}\right]\gamma^{[2]}_m 
=\mathbf{P}^{[2]}_{(n)}\gamma^{[2]}_m$ and imposing Eq.~$(\ref{stable})$
on $\gamma^{[2]}_n $, gives
\begin{equation}
  \sum_{m=1}^{N_0^{[2]}}\Gamma^{[2]}_{nm} u^m_{[2]} =
  -\left\langle \mathbf{d}\gamma^{[2]}_n\Big\vert
  \overline{\mathbf{X}}^{[2]}_{L}\right\rangle,\>\hbox{where }
  \Gamma^{[2]}_{nm} := \left\langle
  \mathbf{d}\gamma^{[2]}_n\Big\vert \mathbf{P}^{[2]}_{(m)}\right\rangle,
\>\> n=1, \dots,
  N_0^{[2]}.
\end{equation}
Then $r^{[2]} := \hbox{rank }\Gamma^{[2]}_{nm}$,
of the remaining $u^m_{[2]}$ functions are determined, up to
$N^{[3]}_0=N_0^{[2]}-r^{[2]}$ \textbf{third-order Lagrangain
  constraint functions},
\begin{equation}
  \gamma^{[3]}_{n_{[3]}} =
  \left\langle\mathbf{d}\gamma^{[2]}_{n_{[3]}}\Big\vert
  \overline{\mathbf{X}}^{[2]}_{L}\right\rangle, \> n_{[3]} = 1, \dots,
  N_0^{[3]},
\end{equation}
are introduced with the conditions $\gamma^{[3]}_{n_{[3]}}=0$ 
imposed. With 
\begin{equation}
I_{[3]} := \hbox{rank } \left\{ 
  \mathbf{d}\gamma_{(n_{[1]})}^{[1]},
 \mathbf{d}\gamma_{(n_{[2]})}^{[2]},
 \mathbf{d}\gamma_{(n_{[3]})}^{[3]}
 \right\},
\end{equation}
independent functions in $\hbox{C}^{[3]}_L :=
\hbox{C}_L^{[2]}\cup\left\{\gamma_{n_{[3]}}^{[3]}, n_{[3]}=1,
\dots, N_0^{[3]}\right\}$, we now have the \textbf{third-order
  constraint submanifold},  
\begin{equation}
  \mathbb{P}_L^{[3]}:=\left\{\mathfrak{u}\in \mathbb{P}^{[2]}_L
  \ \backslash \ \gamma^{[3]}_{n_{[3]}}(\mathfrak{u})=0, n_{[3]}=1,
  \dots, N_0^{[3]}\right\}.
\end{equation}
Once again, the process stops when
$I_{[3]}=I_{[2]}$ or $I_{[3]}=2D$. However, if $I_{[2]}<I_{[3]}<2D$, the
process continues until at the $n_F$-step either
$I_{[n_F]}=I_{[n_F]-1}$ or $I_{[n_F]}=2D$. 

The end result of this algorithm is 
\begin{enumerate}
\item{A submanifold $\mathbb{P}^{[n_F]}_L\subset \mathbb{P}_L$ on which
  dynamics takes place.} 
  \item{A collection
    $\hbox{C}_L^{[n_F]}\subset\overline{\mathcal{F}}$ of constraint
    functions of order $1$ to $n_F$.} 
\item{A second-order, Euler-Lagrange vector field
  \begin{equation}
    \overline{\mathbf{X}}_{EL}^{[n_F]}=\overline{\mathbf{X}}_{L}^{[n_F]} +
    \sum_{m=1}^{N^{[n_F]}_0}u^m_{[n_F]}(\mathfrak{u})\left[\mathbf{P}_{(m)}^{[n_F]}\right],  
  \end{equation}
  with $N_0^{[n_F]}$ arbitrary functions
  $u^m_{[n_F]}(\mathfrak{u})\in\overline{\mathcal{F}}$ for $m=1,
  \dots, N_0^{[n_F]}$, and 
  \begin{equation}
    \overline{\mathbf{X}}_{L}^{[n_F]} =
    \overline{\mathbf{X}}_{L}^{[1]}+
    \sum_{m=N_0^{n_F}+1}^{N_0^{[1]}}u^m_{[n_F]}(\mathfrak{u})
    \left[\mathbf{P}_{(m)}^{[n_F]}\right],  
  \end{equation}
  where the $N_0^{[1]}-N_0^{[n_f]}$ functions
  $u^m_{[n_F]}(\mathfrak{u})\in\overline{\mathcal{F}}$, 
  $m=N_0^{[n_F]}+1, \dots, N_0^{[1]}$, have been uniquely
  determined through the constraint algorithm.} 
\end{enumerate}
Importantly, the end result of the constraint algorithm
$\overline{\mathbf{X}}_{EL}^{[n_F]}$ is still a SOELVF. 

As with the first-order constraint manifold $\mathbb{P}^{[1]}_L$, we
assume that $\mathbb{P}^{[n_F]}_L$ is non-empty. In addition, we
assume that the rank of $\Gamma^{[l]}_{nm}$ is constant on
  $\mathbb{P}_L$ for each $l=1, \dots, n_F$.
  
\section{The passage to Hamiltonian mechanics\label{Passage}}

The question of whether and how dynamical structures on the Lagrangian
phase space are equivalent to such structures on the Hamiltonian phase
space has had a long history \cite{Got1980, Bat1986} (see also
\cite{Car1988b, Bat1987a, Car1987b, Pon1988, Gra1989, Gra1992b}). These
analyses have focused solely on SOLVFs, and often make use of
pullbacks of structures on the Hamiltonian phase space in the
construction of such operators as the evolution operator $K$ and 
the vector field operator $R$ (see \cite{Car1988b, Bat1987a, Car1987b,
  Pon1988, Gra1989, Gra1992b}) which are used to determine the
projectability of Lagrangian constraints and vector fields on
$\mathbf{T}\mathbb{P}_L$, respectively. However, while primary
Hamiltonian constraints play a 
central role in the construction of both operators, the existence of
such constraints is presumed. Moreover, because of the reliance on primary
constraints, a number of subtleties involving first- and second-class
Hamiltonian constraints must be dealt with.

These subtleties and their conclusions, present for
SOLVFs, are not present for SOELVFs. The approach used here focuses
on the symmetry group $\hbox{Gr}_{\mathcal{S}\hbox{ym}}$,
and the geometric structures inherent to almost regular
Lagrangians. The passage from Lagrangian to Hamiltonian mechanics
follows naturally. Much of the content of this section is new.

\subsection{Projectability of functions and vector fields on
  $\mathbb{P}_L$\label{image}} 

The canonical phase space $\mathbb{P}_C:=\mathbf{T}^{*}\mathbb{Q}$
has the cotangent bundle coordinates $\mathfrak{s}=(q,p)$ with
$q\in\mathbb{Q}$ and $p\in\mathbf{T}^{*}_q\mathbb{Q}$. The fiber
derivative is the map $\mathcal{L}:(q,v)\in\mathbb{P}_L \to (q,
p=\partial L/\partial v)\in\mathbb{P}_C$. For regular Lagrangians,
its action on a function $f(\mathfrak{u})\in C^\infty$ on
$\mathbb{P}_L$ gives the function $f_c(\mathfrak{s}) :=
(f\circ\mathcal{L}^{-1})(\mathfrak{s}) =
f(\mathcal{L}^{-1}(\mathfrak{s})) =
f(\mathfrak{u})\vert_{\mathcal{L}(\mathfrak{u})=\mathfrak{s}}$, on
$\mathbb{P}_C$. The action of $\mathcal{L}$ on a vector field
$\mathbf{X} \in\mathbf{T}_{\mathfrak{u}}\mathbb{P}_L$ is given by the
pushforward map $\mathcal{L}_{*}: 
\mathbf{T}_{\mathfrak{u}}\mathbb{P}_L \to 
\mathbf{T}_{\mathcal{L}(\mathfrak{u})}\mathbb{P}_C$, while its action
on a one-form
$\bm{\sigma}\in\mathbf{T}_{\mathfrak{s}}^{*}\mathbb{P}_C$ is given by
the pullback map 
$\mathcal{L}^{*}:\mathbf{T}^{*}_{\mathcal{L}(\mathfrak{u})}\mathbb{P}_C\to
  \mathbf{T}^{*}_{\mathfrak{u}}\mathbb{P}_L$. 
  
The situation changes for singular Lagrangians. While the pullback of
one-forms simply involves replacing 
$\mathfrak{s}$ by $\mathcal{L}(\mathfrak{u})$, the action of
both $\mathcal{L}$ and $\mathcal{L}_{*}$ involve solving for
$\mathfrak{u}$ in $\mathfrak{s}=\mathcal{L}(\mathfrak{u})$.
For singular Lagrangians solutions of this equation are not single
valued, but instead gives the \textbf{preimage of $\mathcal{L}$}, 
\begin{equation}
  \mathcal{L}^{-1}(\mathfrak{s}) :=\left\{\mathfrak{u}\in \mathbb{P}_L
  \ \backslash \ \mathcal{L}(\mathfrak{u}) =\mathfrak{s}\right\},
\end{equation}
which is a submanifold of $\mathbb{P}_L$. As such, the pullback of
a function now results in the collection of functions 
\begin{equation}
  (f\circ\mathcal{L}^{-1})(\mathfrak{s})=\left\{f(\mathfrak{u})
  \ \backslash
  \ \mathfrak{u}\in\mathcal{L}^{-1}(\mathfrak{s})\right\},
  \label{func}
\end{equation}
while the pushforward $\mathcal{L}_{*}\mathbf{X}$ of $\mathbf{X}$
gives the collection of vectors
\begin{equation}
\mathcal{L}_{*}\mathbf{X}(\mathfrak{s}) =
\left\{\mathbf{X}_C(\mathfrak{u}) \ \backslash
\ \mathfrak{u} \in \mathcal{L}^{-1}(\mathfrak{s})
\right\}.
\label{vectors}
\end{equation}
This ambiguity for Eq.~$(\ref{func})$ can be avoided by
focusing on functions that are constant on
$\mathcal{L}^{-1}(\mathfrak{s})$. Then  
$f(\mathfrak{u})=f_C(\mathfrak{s})$ $\forall  
\mathfrak{u}\in\mathcal{L}^{-1}(\mathfrak{s})$, so that 
\begin{equation}
  (f\circ\mathcal{L}^{-1})(\mathfrak{s}) = \left\{f(\mathfrak{u})
  \ \backslash
  \ \mathfrak{u}\in\mathcal{L}^{-1}(\mathfrak{s})\right\}=f_C(\mathfrak{s}),
  \label{Proj-f}
\end{equation}
and is thus single valued. Following the literature, we say that a
function $f$ on $\mathbb{P}_L$ is \textbf{projectable} when
Eq.~(\ref{Proj-f}) holds. It is well known \cite{Got1979,
  Mar1983, Car1988a, Car1990a, Gra1992a, Gra2001} that the condition
for a function $f$ to be projectable is  
\begin{equation}
  \mathbf{G}f(\mathfrak{u})=0,
  \label{Gf}
\end{equation}
for all $\mathbf{G}\in\mathcal{G}$. 

For vector fields, the ambiguity Eq.~$(\ref{vectors})$ is avoided when
the components of $\mathbf{X}_C$ are constant on the preimage. Then
\begin{equation}
\mathcal{L}_{*}\mathbf{X}(\mathfrak{s}) =
\left\{\mathbf{X}_C(\mathfrak{u}) \ \backslash
\ \mathfrak{u} \in \mathcal{L}^{-1}(\mathfrak{s})
\right\}=\mathbf{X}_C(\mathfrak{s}),
\label{Proj-V}
\end{equation}
and we say a vector field on $\mathbf{T}_{\mathfrak{u}}\mathbb{P}_L$
is projectable when Eq.~(\ref{Proj-V}) holds. To determine which
vectors are projectable, consider first the collection of vectors in
$\mathbf{T}_{\mathfrak{u}}\mathbb{P}_L$ for which
$\mathcal{G}$ is an ideal, 
\begin{equation}
\overline{\mathbf{T}_{\mathfrak{u}}\mathbb{P}_L} :=
\left\{\mathbf{X}\in \mathbf{T}_{\mathfrak{u}}\mathbb{P}_L
\ \backslash \ \left[\mathbf{X},\mathbf{G}\right] \in\mathcal{G} \>
\forall\> \mathbf{G}\in\mathcal{G}\right\},
\end{equation}
Applying the same arguments using \textbf{Lemma \ref{basis}} to
$\overline{\mathbf{T}_{\mathfrak{u}}\mathbb{P}_L}$ as was applied to
$\overline{\hbox{ker }\bm{\Omega}_L(\mathfrak{u})}$ gives similar
  results: dim $\overline{\mathbf{T}_{\mathfrak{u}}\mathbb{P}_L}=2D$,
  and $\overline{\mathbf{T}_{\mathfrak{u}}\mathbb{P}_L}$ is
  involutive. The equivalence relation 
  $\overline{\mathbf{X}}_1\sim\overline{\mathbf{X}}_2$ iff
  $\overline{\mathbf{X}}_1-\overline{\mathbf{X}}_2\in\mathcal{G}$
  then follows along with the quotient space
  $\overline{\mathbf{T}_{\mathfrak{u}}\mathbb{P}_L}/\mathcal{G}$. 
\begin{theorem} 
$\overline{\mathbf{T}_{\mathfrak{u}}\mathbb{P}_L}/\mathcal{G}$ 
  is projectable.   
      \label{u-X}

  \begin{proof} Choose an open covering $\mathfrak{U}$ of
    $\mathbb{P}_L$, and a point $\mathfrak{u}$ in an open
    neighborhood $U_{\mathfrak{u}}\in\bm{\mathfrak{U}}$ such that
    $\mathbf{G}\mathfrak{u}^A=0$, $A=1, \dots, 2D$, for all
    $\mathbf{G}\in\mathcal{G}$. Choose also a $\mathbf{X}\in
    \overline{\mathbf{T}_{\mathfrak{u}}\mathbb{P}_L}/\mathcal{G}$, 
    and consider the path $\mathfrak{u}_{\mathbf{X}}(t,\mathfrak{u})$
    given by
    \begin{equation}
      \frac{d\mathfrak{u}_{\mathbf{X}}}{dt} =
        \mathbf{X}\left(\mathfrak{u}_{\mathbf{X}}\right),
      \label{DeQ}
    \end{equation}
    with $\mathfrak{u}_{\mathbf{X}}(0,\mathfrak{u}) =
    \mathfrak{u}$. This open neighborhood can always be chosen small
    enough such that,
    \begin{equation}
      \mathfrak{u}_{\mathbf{X}}^A(t,\mathfrak{u})=e^{t\mathbf{X}}\mathfrak{u}^A,
      \label{exp-X}
    \end{equation}
    on $U_{\mathfrak{u}}$. 
    Then as $\mathcal{G}$ is an ideal of
    $\overline{\mathbf{T}_{\mathfrak{u}}\mathbb{P}_L}$, 
    $e^{-t\mathbf{X}}\mathbf{G}e^{t\mathbf{X}}\in\mathcal{G}$,
    and $\mathbf{G}\mathfrak{u}^A_{\mathbf{X}}(t,\mathfrak{u})=0$ in
    $U_{\mathfrak{u}}$. By applying Eq.~(\ref{exp-X}) to a sequence of
    such open neighborhoods, we can extend this result to any
    connected region $\mathcal{R}$ of $\mathbb{P}_L$. Importantly, as
    the path $\mathfrak{u}_{\mathbf{X}}(t, \mathfrak{u})$ is
    projectable on $\mathcal{R}$, there is the path
    $\mathfrak{s}_{\bm{\mathfrak{X}}}(t, \mathcal{L}(\mathfrak{u})) =
    \mathcal{L}(\mathfrak{u}_{\mathbf{X}}(t, \mathfrak{u}))$ on
    $\mathbb{P}_C$ with tangent vector $\bm{\mathfrak{X}}$ and
    initial data $\mathfrak{s}_{\bm{\mathfrak{X}}}(0) =
    \mathcal{L}(\mathfrak{u})$. The integral flow 
    $\mathfrak{u}_{\mathbf{X}}(t, \mathfrak{u})$ is unique for a 
    given $\mathbf{X}$ and $\mathfrak{u}$. Similarly, the integral
    flow $\mathfrak{s}_{\bm{\mathfrak{X}}}(t,
    \mathcal{L}(\mathfrak{u}))$ is unique for a given
    $\bm{\mathfrak{X}}$ and initial data 
    $\mathfrak{s} = \mathcal{L}(\mathfrak{u})$. As the projection of
    $\mathfrak{u}_{\mathbf{X}}(t)$ to
    $\mathfrak{s}_{\bm{\mathfrak{X}}}(t, \mathcal{L}(\mathfrak{u}))$
    is also unique, we conclude that each $\mathbf{X} \in
    \overline{\mathbf{T}_{\mathfrak{u}}\mathbb{P}_L}/\mathcal{G}$
    is projectable with $\bm{\mathfrak{X}}=\mathcal{L}_{*}\mathbf{X}$.      
  \end{proof}
  \label{IVector}
\end{theorem}
(A coordinate-based proof using Eq.~$(\ref{Gf})$ can also be given.) The
converse is also true, as we show in the next section. 

\subsection{Projection of dynamical structures}

By construction, 
$\overline{\mathcal{F}}$ is projectable, and as both
$u^m(\mathfrak{u})$ and $\gamma\in\overline{\mathcal{F}}$ for any
$\gamma\in\hbox{C}_L^{[n_F]}$, they also are projectable. In
addition, $\mathbf{G}E =0$, and $E$ is projectable with
its image $H_C=(E\circ\mathcal{L}^{-1})(\mathfrak{s})$ being the canonical
Hamiltonian. With the exception of the energy, we avoid introducing
new notation, and will represent the projection of any 
function $f(\mathfrak{u})\in \overline{\mathcal{F}}$ through its
argument: $f(\mathfrak{s})$. 

Both $\overline{\mathcal{S}\hbox{ol}}$ and
$\overline{\hbox{ker }\mathbf{\Omega}_L(\mathfrak{u})}/\mathcal{G}$
are subsets of
$\overline{\mathbf{T}_{\mathfrak{u}}\mathbb{P}_L/\mathcal{G}}$, and
are projectable. Of particular interest are  
\begin{equation} 
  \hbox{Prim}
  :=\mathcal{L}_{*}\left(\overline{\hbox{ker
    }\mathbf{\Omega}_L(\mathfrak{u})}\right)=
  \{\mathfrak{P}\in\mathbf{T}_{\mathfrak{s}}\mathbb{P}_C 
  \ \backslash \ \mathfrak{P} = \mathcal{L}_{*} [\mathbf{P}],
  \> \forall\> [\mathbf{P}]\in\overline{\hbox{ker 
    }\bm{\Omega}_L(\mathfrak{u})}/\mathcal{G}\},
\end{equation}
and
\begin{equation}
  \hbox{Flow}_{H_T}
  :=\mathcal{L}_{*}\left(\overline{\mathcal{S}\hbox{ol}}\right)=
  \big\{\mathfrak{X}_{H_T}\in 
  \mathbf{T}_{\mathfrak{s}=\mathcal{L}(\mathfrak{u})}\mathbb{P}_C \ \backslash
  \ \mathfrak{X}_{H_T} = \mathcal{L}_{*}\overline{\mathbf{X}}_{EL}
  \> \forall \> 
  \overline{\mathbf{X}}_{EL}\in\overline{\mathcal{S}\hbox{ol}}\big\}. 
\end{equation}
In particular, the general $\overline{\mathbf{X}}_{EL}$ in
Eq.~$(\ref{EL})$ gives the general vector field 
\begin{equation}
  \mathfrak{X}_{H_T} = \overline{X}_L^{qa}(\mathfrak{s})\frac{\bm{\partial}
    \>\>\>}{\bm{\partial} q^a} +
  \left[\overline{X}_L^{qa}N_{ab}\right]\big\vert_{\mathfrak{s}}
  \frac{\bm{\partial}\>\>\>}{\bm{\partial} p_b} - \frac{\partial H_C}{\partial
    q^b}\frac{\bm{\partial} \>\>\>}{\bm{\partial} p_b}+
  \sum_{m=1}^{N_0}
  u^m(\mathfrak{s})\mathfrak{P}_{(m)},
  \label{GenHE}
\end{equation}
in $\hbox{Flow}_{H_T}$ when expressed in terms of local coordinates. Here,
\begin{equation}
  N_{ab} = \frac{\partial^2 L}{\partial v^a\partial q^b},
\end{equation}
and $\mathfrak{P}_{(m)} =
\mathcal{L}_{*}[\mathbf{P}_{(m)}]$ for a choice $\{[\mathbf{P}_{(m)}], m=1,
\dots, N_0\}$ of basis for $\overline{\hbox{ker
  }\bm{\Omega}_L(\mathfrak{u})}/\mathcal{G}$. 

With the canonical two-form $\bm{\omega}
= \bm{d}q^a\wedge \bm{d}p_a$ on
$\mathbf{T}^{*}_{\mathfrak{s}}\mathbb{P}_C$, we have the collection
of one-forms,
\begin{equation}
  \hbox{Prim}^{\flat} := \left\{\bm{\pi}\in
  \bm{\Lambda}^1(\mathcal{L}(\mathbb{P}_L)) 
  \ \backslash \ \bm{\pi} = \omega^\flat\mathfrak{P}\> \forall\>
  \mathfrak{P}\in\hbox{Prim}\right\}.
\end{equation}
which gives the primary constraints, and 
\begin{equation}
  \hbox{Flow}_{H_T}^{\flat} := \left\{\bm{\alpha}\in
  \bm{\Lambda}^1(\mathcal{L}(\mathbb{P}_L)) 
  \ \backslash \ \bm{\alpha} = \omega^\flat \mathfrak{X}_{H_T}\> \forall\>
  \mathfrak{X}_{H_T}\in\hbox{Flow}_{H_T}\right\},
\end{equation}
which gives the set of total Hamiltonians.

\subsubsection{$\bm{\hbox{Prim}}$ and the Primary Hamiltonian Constraints}

Using the kernel of the pullback map,
\begin{equation}
  \hbox{ker }\mathcal{L}^{*}
  :=\left\{\bm{\phi}\in\bm{\Lambda}(\mathbb{P}_C)
  \ \backslash 
  \ \mathcal{L}^{*}\bm{\phi} =0\right\},
\end{equation}
in this section we construct from 
$\overline{\hbox{ker
  }\mathbf{\Omega}_L(\mathfrak{u})}/\mathcal{G}$
the primary Hamiltonian constraints.  

\begin{lemma}
  \label{kerLstar}
  For any one-form $\bm{\sigma}\in\bm{\Lambda}^1(\mathcal{L}(\mathbb{P}_L))$,
  $\bm{\sigma} \in\hbox{ker }\mathcal{L}^{*}$ iff $\bm{\sigma} \in
  \hbox{Prim}^\flat$. 

  \begin{proof}
    Suppose first that $\bm{\sigma}\in \hbox{Prim}^\flat$. Then there
    exists a $[\mathbf{P}]\in\overline{\hbox{ker
      }\mathbf{\Omega}_L(\mathfrak{u})}/\mathcal{G}$ such that
    $\bm{\sigma}=i_{\mathcal{L}_{*}[\mathbf{P}]}\bm{\omega}$. As
    $\mathcal{L}^{*}\bm{\omega} =\mathbf{\Omega}_L$,   
    $\mathcal{L}^{*}\bm{\sigma}=i_{\mathbf{P}}\mathbf{\Omega}_L=0$,
    and it follows that $\bm{\sigma}\in \hbox{ker }\mathcal{L}^{*}$.

    Next suppose that $\bm{\sigma}\in\hbox{ker
    }\mathcal{L}^{*}$. Let $\mathfrak{X}$ be the unique vector in
    $\mathbf{T}_{\mathfrak{s}}\mathbb{P}_C$ such that
    $i_{\mathfrak{X}}\bm{\omega} = \bm{\sigma}$. Then $\mathcal{L}^{*}\left[i_{\mathfrak{X}}\bm{\omega}\right]=0$.
    But as both $i_{\mathfrak{X}}\bm{\omega}$ and $\bm{\omega}$ are
    differential forms, their pullbacks are well-defined and there
    must then be a $\mathbf{X}\in\mathbf{T}_{\mathbf{u}}\mathbb{P}_L$
    such that $\mathcal{L}_{*}\mathbf{X}=\mathfrak{X}$. It then
    follows that $i_{\mathbf{X}}\bm{\Omega}_L=0$, and
    thus $\bm{\sigma}\in\hbox{Prim}^\flat$.   
  \end{proof}
\end{lemma}
(A coordinate-based proof of this lemma can also be given.)

Consider now the Pfaff system of exterior
equations,  
\begin{equation}
  {Pf}(\hbox{Prim}^\flat) := \left\{\bm{\pi}_{(n)}=0, n=1, \dots,
  N_0 \right\}. 
\end{equation}
and the \textbf{integral manifold} $(\mathbb{P}_L, \mathcal{L})$
of ${Pf}(\hbox{Prim}^\flat)$ \cite{Cho1982}. As
$N_0=\hbox{dim }\left(\overline{\hbox{ker
    }\bm{\Omega}_L(\mathfrak{u})}/\mathcal{G}\right)=  
  \hbox{dim }\hbox{Prim}=\hbox{dim }\hbox{Prim}^\flat$,
  $\hbox{rank }Pf(\hbox{Prim}^\flat)=N_0$.  
Of particular interest is the ideal \cite{Cho1982} of
$Pf(\hbox{Prim}^\flat)$  
\begin{equation}
  I[Pf(\hbox{Prim}^\flat)] := \Bigg\{\sum_{n=1}^{N_0}
  \bm{\xi}^n\wedge\bm{\pi}_{(n)} \ \backslash
    \ \bm{\xi}^n\in\bm{\Lambda}(\mathbb{P}_C),   \bm{\pi}_{(n)} \in Pf(\hbox{Prim}^\flat)\Bigg\}.
\end{equation}

\begin{lemma}
  \label{ideal}
  $I[Pf(\hbox{Prim}^\flat)] = \hbox{ker }\mathcal{L}^{*}$.

  \begin{proof} If $\bm{\sigma}\in I[Pf(\hbox{Prim}^\flat)]$, then
    \begin{equation}
      \bm{\sigma} = \sum_{n=1}^{N_0} \bm{\xi}^n\wedge\bm{\pi}_{(n)}.
    \end{equation}
    From \textbf{Lemma \ref{kerLstar}}, 
    \begin{equation}
      \mathcal{L}^{*}\bm{\sigma} = \sum_{n=1}^{N_0}
      \mathcal{L}^{*}\bm{\xi}^n\wedge\mathcal{L}^{*}\bm{\pi}_{(n)} = 0,
    \end{equation}
    so that $I[Pf(\hbox{Prim}^\flat)] \subseteq \hbox{ker
    }\mathcal{L}^{*}$.

    Next, choose a basis $\bm{\theta}_{(n)}, n=1, \dots, 2D$ of
    $\bm{\Lambda}^1(\mathcal{L}(\mathbb{P}_L))$ such that
    $\bm{\theta}_{(n)} = \bm{\pi}_{(n)}$ for $n=1, \dots, N_0$. Let
    $\bm{\sigma}\in\hbox{ker }\mathcal{L}^{*}$ be the $p$-form,
    \begin{equation}
      \bm{\sigma}(\mathfrak{s}):=\frac{1}{p!} \sum_{n_1, \dots n_p=1}^D\sigma_{n_1\dots n_p}(\mathfrak{s})
      \bm{\theta}_{(n_1)}\wedge\dots\wedge \bm{\theta}_{(n_p)}.
    \end{equation}
    Then as $\mathcal{L}^{*}\bm{\sigma}=0$, 
    \begin{equation}
      0= \frac{1}{p!} \sum_{n_1, \dots,
        n_p=1}^D\sigma_{n_1\dots n_p}(\mathcal{L}(\mathfrak{u}))
      \mathcal{L}^{*}\bm{\theta}_{(n_1)}\wedge\dots\wedge
      \mathcal{L}^{*}\bm{\theta}_{(n_p)},
    \end{equation}
    and from \textbf{Lemma \ref{kerLstar}} we conclude
    that $\sigma_{n_1 \dots n_p}(\mathcal{L}(\mathfrak{u}))=0$ for
    $n_s>N_0$, $s=1, \dots, p$. Thus there exists forms
    $\bm{\xi}^{(n)}$ such that 
    \begin{equation}
      \bm{\sigma}=\sum_{n=1}^{N_0} \bm{\xi}^{(n)}\wedge\bm{\pi}_{(n)},
    \end{equation}
    so that $\hbox{ker }\mathcal{L}^{*}\subseteq I[Pf(\hbox{Prim}^\flat)]$
    as well. 
    \end{proof}
\end{lemma}
The construction of the primary Hamiltonian constraints is now
trivial.

Consider a $\bm{\pi}\in Pf(\hbox{Prim}^\flat)$. As
$\mathcal{L}^{*}\bm{d}\bm{\pi} = \bm{d}\mathcal{L}^{*}\bm{\pi} =0$,
$\bm{d\pi}\in \hbox{ker } \mathcal{L}_{*}$, and from \textbf{Lemma
  \ref{ideal}}, $\bm{d\pi}\in
I[Pf(\hbox{Prim}^\flat)]$. There are then
one-forms $\bm{\xi}^{(n)}$, $n=1, \dots, N_0$ such that  
      \begin{equation}
        \bm{d\pi} = \sum_{n=1}^{N_0}\bm{\xi}^{(n)}\wedge\bm{\pi}_{(n)}.
      \end{equation}
Then $\bm{d\pi}\wedge\bm{\pi}_{(1)}\wedge \dots\wedge
\bm{\pi}_{(N_0)}=0$, and thus
$Pf(\hbox{Prim}^\flat)$ is closed
\cite{Cho1982}. It follows from the Frobenius theorem that
$Pf(\hbox{Prim}^\flat)$ is completely 
integrable. There are then $N_0$ first integrals $\gamma^{[0]}_n$ of
$Pf(\hbox{Prim}^\flat)$ such that in 
a neighborhood about each generic point
$\mathfrak{u}\in\mathbb{P}_C$, $\{\bm{\pi}_{(n)}=0\}\sim
\{\bm{d}\gamma^{[0]}_n=0\}$; these forms may be chosen such 
that $\bm{\pi}_{(n)} = f_n(\mathfrak{s}) \bm{d}\gamma^{[0]}_n$ where $f_n$
is a $C^\infty$ function on $\mathbb{P}_C$. The functions $\gamma^{[0]}_n$
are the primary Hamiltonian constraints while 
\begin{equation}
  \mathbb{P}^{[0]}_C:=\left\{\mathfrak{s}\in\mathbb{P}_C
  \ \backslash \ \gamma^{[0]}_n(\mathfrak{s})=0, n = 1, \cdots,
  N_0\right\}, 
\end{equation}
is the \textbf{primary constraint submanifold}. Connections between
the primary constraints and vectors in ker
$\mathbf{\Omega}_L(\mathfrak{u})$ have been found previously
by using the time-evolution operator $K$ \cite{Gra2001}. Such analyses
make use of pullbacks of the primary Hamiltonian constraints, however, while the
approach here is constructive.

\begin{lemma} $\mathcal{L}_{*}\left(
  \overline{\mathbf{T}_{\mathfrak{u}}\mathbb{P}_L}/\mathcal{G}\right)
  =
  \mathbf{T}_{\mathfrak{s}=\mathcal{L}(\mathfrak{u})}\mathbb{P}^{[0]}_C$.

  \begin{proof} Let $[\overline{\mathbf{X}}]\in
    \overline{\mathbf{T}_{\mathfrak{u}}\mathbb{P}_L}/\mathcal{G}$. As
    $[\overline{\mathbf{X}}]$ is projectable, $\langle
    \bm{d} \gamma^{[0]}_n\vert
    \mathcal{L}_{*}[\overline{\mathbf{X}}]\rangle = \langle
    \mathcal{L}^{*}\bm{d} \gamma^{[0]}_n\vert 
    [\overline{\mathbf{X}}]\rangle=0$ since $\bm{d}\gamma^{[0]}_n\in
    \hbox{ker }\mathcal{L}^{*}$, and it follows that $\mathcal{L}_{*}\left(
  \overline{\mathbf{T}_{\mathfrak{u}}\mathbb{P}_L}/\mathcal{G}\right)
  \subseteq
  \mathbf{T}_{\mathfrak{s}=\mathcal{L}(\mathfrak{u})}\mathbb{P}^{[0]}_C$. But as
  dim
  $\overline{\mathbf{T}_{\mathfrak{u}}\mathbb{P}_L}/\mathcal{G}]  
  = 2D-N_0 = \hbox{dim }
  \mathbf{T}_{\mathfrak{s}=\mathcal{L}(\mathfrak{u})}\mathbb{P}^{[0]}_C$,
    $\mathcal{L}_{*}\left(\overline{\mathbf{T}_{\mathfrak{u}}\mathbb{P}_L}/
  \mathcal{G}\right) =
  \mathbf{T}_{\mathfrak{s}=\mathcal{L}(\mathfrak{u})}\mathbb{P}^{[0]}_C$
  follows.  
  \end{proof}
  \label{T-P}
\end{lemma}
The converse of \textbf{Theorem \ref{u-X}} then follows. Importantly,
because
$\mathcal{L}_{*}\left(\overline{\mathcal{S}\hbox{ol}}\right)\subset 
\mathbf{T}_{\mathfrak{s}=\mathcal{L}(\mathfrak{u})}\mathbb{P}^{[0]}_C$,
the integral flow fields of SOELVFs lie on $\mathbb{P}^{[0]}_C$.

\subsubsection{$\overline{\mathcal{S}\hbox{ol}}$ and the total Hamiltonian}  

On $\mathbb{P}^{[1]}_L$, $\bm{\beta}[\mathbf{X}_E]=0$, and the energy
equation may be written as
$0=\bm{d}E-i_{\overline{\mathbf{X}}_L}\mathcal{L}^{*}\bm{\omega}$. It
follows that  
\begin{equation}
  i_{\mathcal{L}_{*}\overline{\mathbf{X}}_L}\bm{\omega} =\bm{d}H_C,
\end{equation}
from which we conclude that if $\bm{\mathfrak{X}}_C$ is the
Hamiltonian flow field for $H_C$, then $\mathfrak{X}_C =  
\mathcal{L}_{*}\overline{\mathbf{X}}_L$.
The image of the pushforward of Eq.~$(\ref{EL})$ gives the vector field
$\mathfrak{X}_{H_T} := 
\mathcal{L}_{*}\overline{\mathbf{X}}_{EL} \in
\hbox{Flow}_{H_T}$, 
\begin{equation}
  \mathfrak{X}_{H_T} 
  =\mathfrak{X}_C +
  \sum_{m=1}^{N_0}u^m(\mathfrak{s})\mathfrak{P}_{(m)},
  \label{XHE}
\end{equation}
that is everywhere tangent to $\mathbb{P}_C^{[0]}$. Correspondingly, a
general one form in $\hbox{Flow}_{H_T}^\flat$ is  
\begin{equation}
  i_{\mathfrak{X}_{H_T}}\bm{\omega} = \bm{d}H_C+\sum_{m=1}^{N_0}
    u^m(\mathfrak{s})f_m(\mathfrak{s}) \bm{d\gamma}^{[0]}_{m},
\end{equation}
which gives the total Hamiltonian,
\begin{equation}
  H_T = H_C+\sum_{m=1}^{N_0}
  u^m(\mathfrak{s})f_n(\mathfrak{s}) \bm{\gamma}^{[0]}_{m},
  \label{HE}
\end{equation}
for the dynamical system. This leads to the sequence of maps: 
\begin{equation}
  \overline{\mathcal{S}\hbox{ol}}\quad {\mathfrak{L}_{*} \atop
    \xrightarrow{\hspace*{1.5cm}}} \quad \hbox{Flow}_{H_T}\quad{\omega^\flat \atop
    \xrightarrow{\hspace*{1.5cm}}}\quad \hbox{Flow}_{H_T}^\flat \quad {\mathfrak{L}^{*}
    \atop \xrightarrow{\hspace*{1.5cm}}}\quad E, 
\end{equation}
and to each $\overline{\mathbf{X}}_{EL}\in\overline{\mathcal{S}\hbox{ol}}$
there is a corresponding total Hamiltonian
$H_T\in\hbox{Flow}_{H_T}^\flat$. 

\subsubsection{The Equivalence of the Constraint Algorithm for Lagrangians
  and the Stability Analysis of Canonical Hamiltonians} 

It is well known that the integral flow generated by
$\mathfrak{X}_{H_T}$ need not be confined to $\mathbb{P}_C^{[0]}$
even though its initial data is chosen to be on this 
submanifold. This difficulty is resolved through a 
\textbf{stability analysis} \cite{Hen1992} where 
$\{H_T,\bm{\gamma}^{[0]}_n\}=0$ is imposed on the primary constraints,
and when necessary, successively on the secondary, tertiary, and
higher-level Hamiltonian constraints. While this process is traditionally
applied to the canonical Hamiltonian, \textbf{Section \ref{&StabC}}
describes a constraint algorithm for SOELVFs. We show 
here that this constraint algorithm is equivalent 
to the stability analysis of the canonical Hamiltonian.  

Choose a $\overline{\mathbf{X}}_{EL}^{[1]}\in
\overline{\mathcal{S}\hbox{ol}}$, where 
we follow the notation established in Eq.~$(\ref{notation})$. There
is then a corresponding $\mathfrak{X}_{H_T}^{[1]} =
\mathfrak{L}_{*}\overline{\mathbf{X}}_{EL}^{[1]}$, and total Hamiltonian
$H_T^{[1]}$. The stability analysis of the primary constraints under
$H_T^{[1]}$ then results in
\begin{equation}
  f_n\frac{d\gamma^{[0]}_n}{dt}
    = \left\langle \bm{\pi}_{(n)}\vert \mathfrak{X}_{C}\right\rangle
    +
    \sum_{m=1}^{N_0} u^m \left\langle \bm{\pi}_{(n)}\vert
    \mathfrak{P}_{(m)}\right\rangle,  
\end{equation}
after using Eq.~$(\ref{XHE})$. But $\left\langle \bm{\pi}_{(n)}\vert
\mathfrak{P}_{(m)}\right\rangle= \langle\bm{\omega}\vert
\mathcal{L}_{*}(\mathbf{P}_{(n)})\otimes
\mathcal{L}_{*}(\mathbf{P}_{(m)})\rangle = \langle \mathbf{\Omega}_L\vert
\mathbf{P}_{(n)}\otimes\mathbf{P}_{(m)}\rangle =0$, while $
\langle \bm{\pi}_{(n)}\vert \mathfrak{X}_{C}\rangle=
\langle\bm{\omega}\vert
\mathcal{L}_{*}(\mathbf{P}_{(n)})\otimes\mathcal{L}_{*}\left(\overline{\mathbf{X}}_L\right)
\rangle 
=
\langle\mathbf{\Omega}_L\vert
\mathbf{P}_{(n)}\otimes\overline{\mathbf{X}}_L \rangle
=-\langle\mathbf{d}E\vert \mathbf{P}_{(n)}\rangle$ since we are on the
$\bm{\beta}[\mathbf{X}_E]=0$ surface. As $\langle\mathbf{d}E\vert
\mathbf{P}_{(n)}\rangle=\gamma^{[1]}_n$, 
\begin{equation}
  f_n\frac{d\gamma^{[0]}_n}{dt} = -\gamma_{n}^{[1]}(\mathfrak{s}).
\end{equation}
The projection of first-order constraints automatically gives the
secondary Hamiltonian constraints. It follows that 
$\{H_T,\bm{\gamma}^{[0]}_n\}=0$ is automatically satisfied through the
Lagrangian constraint condition $\gamma^{[1]}_n(\mathfrak{u})=0$.

The stability analysis must now be applied to the secondary
constraints:
$\mathcal{L}_{\mathfrak{X}^{[1]}_{H_T}}\gamma^{[1]}_n(\mathfrak{s})=0$. But
as $\mathfrak{X}^{[1]}_{H_T}
=\mathfrak{L}_{*}\overline{\mathbf{X}}_{EL}^{[1]}$, this requirement
is equivalent to imposing the constraint condition:
$\mathcal{L}_{\overline{\mathbf{X}}^{[1]}_{EL}}\gamma^{[1]}_n(\mathfrak{u})=0$. From
\textbf{Section \ref{&StabC}} doing so results in the SOELVF
$\overline{\mathbf{X}}^{[2]}_{EL}$, and thus gives a corresponding
Hamiltonian flow field $\mathfrak{X}^{[2]}_{H_T}$ and total
Hamiltonian $H^{[2]}_T$. If second-order Lagrangian constraints are
introduced at this step, their projection will give the 
tertiary Hamiltonian constraints.  

This procession continues with the stability analysis of the
$n^{\textit{th}}$-level Hamiltonian constraints giving a
$\overline{\mathbf{X}}_{EL}^{[n]}$, and thus a corresponding Hamiltonian
flow field $\mathfrak{X}^{[n]}_{H_T}
=\mathfrak{L}_{*}\overline{\mathbf{X}}_{EL}^{[n]}$ and total
Hamiltonian $H^{[n]}_T$. If $(n+1)^{th}$-level Hamiltonian constraints
are introduced, they are the projection of the $n^{th}$-order Lagrangian
constraints. The analysis stops when the Lagrangian constraint
algorithm ends: at the $n_F$-step. The end result  
$\overline{\mathbf{X}}^{[n_F]}_{EL}$ of the constraint algorithm gives
a $\mathfrak{X}^{[n_F]}_{H_T}$ with integral flows that lie on the
Hamiltonian constraint submanifolds. Correspondingly, there is a $H_T^{[n_F]}$
that agrees with the end result of the stability analysis of the total
Hamiltonian. The Lagrangian constraint algorithm applied to
$\overline{\mathbf{X}}_{EL}$ is thus equivalent to 
the stability analysis of the canonical Hamiltonian.  

\section{Examples of almost regular Lagrangians\label{Exam}}

In this section we present three examples of dynamical systems with
almost regular Lagrangians. The first example describes a 
single particle interacting with an external potential. It
illustrates the role $\mathcal{G}$ plays, and the tight 
relationship between $\hbox{Gr}_{\mathcal{S}\hbox{ym}}$, the symmetries of 
the Euler-Lagrange equations of motion, and the gauge symmetries
of the Lagrangian. Moreover, it explicitly shows that $\mathcal{G}$
is not the generator of the local gauge symmetry, as is sometimes
asserted in the literature. The second example consists of two
interacting particles with a Lagrangian that has 
a local conformal symmetry. It is an example of a dynamical
system for which only a subset of vectors in $\overline{\hbox{ker
  }\mathbf{\Omega}_L(\mathfrak{u})}$ generate the symmetry
group. The third example consists 
of a particle with both local conformal symmetry and
time-reparametization invariance. It is an example of a fully
constrained dynamical
system\textemdash as such, $\mathcal{S}\hbox{ol} = \hbox{ker
}\mathbf{\Omega}_L(\mathfrak{u})$\textemdash that has two gauge
symmetries. The analysis of all three systems are done using the
techniques and tools presented above.  

\subsection{A Lagrangian With and Without a Local Gauge Symmetry}  

Whether the action
\begin{equation}
  S_{1}:=\int\left[\frac{1}{2}m\left(\frac{d\widehat{q}}{dt}\right)^2
    -V(q^a)\right]dt, 
\end{equation}
with $\vert q\vert=\sqrt{q^a q_a}$ and $\widehat{q}^a := q^a/\vert
q\vert$, $a=1, \dots, D$, has a local gauge symmetry depends on the
choice of potential $V(q)$. With one choice both the Lagrangian and
the equations of 
motion have a local gauge symmetry; with another choice the equations
of motion has a symmetry while the Lagrangian
does not have a local gauge symmetry; and with a third choice, neither
has a symmetry. Interestingly, $L$ is 
singular irrespective the choice of $V(q)$, showing that not all
singular Lagrangians need have a symmetry.   

With $\Pi_{ab}(q):= \delta_{ab} - \widehat{q}_a\widehat{q}_b$,
\begin{equation}
  \mathbf{\Omega}_M = \frac{m}{\vert q\vert^2}\Pi_{ab}(q)
  \mathbf{d}q^a\wedge\mathbf{d}v^b,\qquad
  \mathbf{\Omega}_F=\frac{m}{\vert q\vert^3}
  \left(\widehat{q}\cdot\mathbf{d}q\right)\wedge
  \left(v\cdot\Pi(q)\cdot\mathbf{d}q\right),
\end{equation}
and $\mathcal{C}$ and $\mathcal{G}$ are spanned by $\mathbf{U}^q_{(1)}
= \widehat{q}\cdot\bm{\partial}/\bm{\partial} q$ and  
  $\mathbf{U}^v_{(1)} =
  \widehat{q}\cdot\bm{\partial}/\bm{\partial} v$, 
respectively, while $\overline{\hbox{ker
  }\mathbf{\Omega}_L(\mathfrak{u})}$ is spanned by
$\mathbf{U}^q_{(1)}$ and  
\begin{equation}
  P_{(1)}=\widehat{q}\cdot\frac{\bm{\partial}\>\>\>}{\bm{\partial} q} +
    \frac{1}{\vert q\vert}v\cdot\frac{\bm{\partial}\>\>\>}{\bm{\partial}
      v}.
\end{equation}
The energy is
\begin{equation}
  E=\frac{1}{2}\frac{m}{\vert q\vert^2}v\cdot\Pi(q)\cdot v + V(q),
  \label{energy1}
\end{equation}
and there is only one first-order, Lagrangian constraint,
\begin{equation}
  \gamma^{[1]}=\mathbf{U}^q_{(1)} V, 
\end{equation}
with $\bm{\beta}[\mathbf{X}_E] =
\gamma^{[1]}\mathbf{\Theta}^{(1)}_q$, $\mathbf{\Theta}^{(1)}_q =
\widehat{q}\cdot \bm{d} q$. As expected, 
$\mathfrak{L}_{\mathbf{G}}\gamma^{[1]}=0$.

We may choose
\begin{equation}
  \mathbf{X}_{L} =
  v\cdot\frac{\bm{\partial}\>\>\>}{\bm{\partial}q} +
  2\frac{(\widehat{q}\cdot v)}{\vert q\vert}
  v\cdot\frac{\bm{\partial}\>\>\>}{\bm{\partial} v} -
  \frac{\vert q\vert^2}{m} \frac{\partial V}{\partial
    q}\cdot\frac{\bm{\partial}\>\>\>}{\bm{\partial} v}.
\end{equation}
As $[\mathbf{X}_{L}, \mathbf{U}^v_{(1)}] \sim -\mathbf{P}_{(1)}$, a
symmetry transformation of $\mathbf{X}_L$ does not result in a
SOLVF. Instead,  
\begin{equation}
  \overline{\mathbf{X}}_{L} = v\cdot{\Pi(q)}\cdot
  \frac{\bm{\partial}\>\>\>}{\bm{\partial} q} +
  \frac{(\widehat{q}\cdot v)}{\vert q\vert}
  v\cdot\frac{\bm{\partial}\>\>\>}{\partial v} -
  \frac{\vert q\vert^2}{m}\frac{\partial V}{\partial
    q}\cdot\Pi(q)\cdot\frac{\bm{\partial}\>\>\>}{\bm{\partial}v},
\end{equation}
is constructed, and a general SOELVF is 
$\overline{\mathbf{X}}_{EL}=\overline{\mathbf{X}}_{L} + u(\mathfrak{u})
\left[\mathbf{P}_{(1)}\right]$, where
$u(\mathfrak{u})\in\overline{\mathcal{F}}$.
Because
\begin{equation}
  \mathfrak{L}_{\mathbf{P}_{(1)}} \bm{\beta} =
  \mathbf{d}\left[\mathbf{U}^q_{(1)}V
    \right]-\frac{1}{\vert
    q\vert^2}\widehat{q}\cdot\frac{\partial\>\>\>}{\partial 
    q}\left(\Pi_a^{\>\>b}(q)\frac{\partial V}{\partial\widehat{q}^b}\right)
  \mathbf{d}q^a,
  \label{beta}
\end{equation}
whether or not $\mathcal{S}\hbox{ym}$ is empty depends on the
symmetries of $V(q)$. As the constraint 
algorithm gives 
\begin{equation}
  \mathfrak{L}_{\overline{\mathbf{X}}_{EL}}\gamma^{[1]} = v\cdot\Pi
  \cdot\frac{\partial \gamma^{[1]}}{\partial q} + u(\mathfrak{u})
  \mathbf{U}^q_{(1)}\gamma^{[1]},
  \label{stab}
\end{equation}
whether or not $u(\mathfrak{u})$ is determined
also depends on the symmetries of $V(q)$. There are three possibilities,
none of which will require the introduction of higher-order Lagrangian
constraints.   

\bigskip
\noindent{\textit{The symmetric potential}}
\bigskip
    
For $\mathbf{P}_{(1)}$ to generate a symmetry, 
\begin{equation}
  0=\frac{1}{\vert q\vert^2} \widehat{q}\cdot\frac{\partial \>\>\>}{\partial q} \left(\Pi_a^{\>\>b}(q)\frac{\partial
    V}{\partial \widehat{q}^b}\right),
\end{equation}
and it follows that 
\begin{equation}
  \frac{\partial V}{\partial\widehat{q}^a} = \frac{\partial
    V_{AS}(\widehat{q}^a)}{\partial \widehat{q}^a},
\end{equation}
where $V_{AS}$ is a function of $\widehat{q}^a$ only. Then
$\mathbf{P}_{(1)}$ generates a symmetry iff $V(q,\widehat{q}^q) =
V_{Sph}(q)+V_{AS}(\widehat{q}^a)$, where $V_{Sph}$ is a function of
$q$ only. The group $\mathcal{S}\hbox{ym}$ is one-dimensional, and
spanned by $\mathbf{P}_{(1)}$. 

The constraint condition Eq.~$(\ref{stab})$ for this potential
reduces to 
\begin{equation}
  0=u(\mathfrak{u})\frac{d^2V_{Sph}(q)}{d\vert q\vert^2},
\end{equation}
which must be satisfied on $\mathbb{P}_L$. There are two cases:

\bigskip
\textit{Case 1:} $\frac{d^2V_{Sph}}{d\vert q\vert^2}=0$.
\bigskip

\noindent Then $V_{Sph} = aq+b$, but since 
\begin{equation}
  \gamma^{[1]}=\frac{dV_{Sph}}{d\vert q\vert} =a,
\end{equation}
the condition $\gamma^{[1]}=0$ requires $a=0$. As we may choose $b=0$,
$V(q) = V_{AS}(\widehat{q}^a)$ only. The Lagrangian is
invariant under the local conformal transformation $q^a\to \alpha q^a$,
where $\alpha$ is an arbitrary, nonvanishing function on
$\mathbb{P}_L$. The function $u(\mathfrak{u})$ is not determined, and
correspondingly, the dynamics of the 
particle is determined only up to an arbitrary function.

\bigskip
\textit{Case 2:} $\frac{d^2V_{Sph}}{d\vert q\vert^2}\ne0$.
\bigskip

\noindent In this case $u(\mathfrak{u})=0$, and the dynamics of the
particle is completely determined by its initial data. The
Lagrangian does not have a local gauge symmetry. The   
first-order, Lagrangian constraint $\gamma^{[1]}=0$ defines a surface
on $\mathbf{P}_L$, and for dynamics to be possible 
the set of solutions 
\begin{equation}
    \left\{R_i\in\mathbb{R} \ \backslash
    \ \frac{dV_{Sph}}{d\vert q\vert}\Bigg\vert_{R_i} =0\right\},
\end{equation}
must be non-empty. Dynamics are on the surfaces $\vert q\vert
-R_i =0$, and on them the potential reduces to $V(q)=V_{Sph}(R_i) +
V_{AS}(\widehat{q}^a)$. This reduced potential has the same symmetry
as the potential $V_{AS}(\widehat{q}^a)$ in \textit{Case 1}, leading to
equations of motion that have the same generalized Lie symmetry. The
Lagrangian for the two cases, however, do not have the same 
invariances, resulting in one case to dynamics that are determined up to 
an arbitrary $u(\mathfrak{u})$ while in the other case to
$u(\mathfrak{u})=0$ and dynamics
that are completely determined by the choice of initial data. 

A specific example of this type of potential is the Mexican hat
potential: $V(q) = -\lambda \vert q\vert^2/2 +\beta \vert
q\vert^4/4$. Then
\begin{equation}
  \gamma^{[1]}=-\lambda\vert q\vert + \beta \vert q\vert^3.
\end{equation}
As $\vert q\vert\ne 0$, dynamics are thus on the surface $\vert q\vert
= (\beta/\lambda)^{1/2}$ for $\beta/\lambda>0$. This breaks the local
conformal symmetry while preserving rotational symmetry.

\bigskip
\noindent{\textit{The asymmetric potential}}
\bigskip

For a general $V$, the second term in
Eq.~$(\ref{beta})$ does not vanish, $\mathbf{P}_{(1)}$ does not
generate a symmetry of the equations of motion,
$\mathcal{S}\hbox{ym}=\{\emptyset\}$, and Eq.~(\ref{stab}) gives 
\begin{equation}
  u=-\frac{v\cdot\Pi
  \cdot\frac{\partial \gamma^{[1]}}{\partial q}}
  {\mathbf{U}^q_{(1)}\gamma^{[1]}}.
\end{equation}
The dynamics of the particle is uniquely determined by its initial
data.

The passage to Hamiltonian mechanics is straightforward. With the
canonical momentum, $p_a=m\Pi_{ab}(q)v^b/q^2$, Eq.~$(\ref{energy1})$
gives $H_C=q^2p^2/2m+V(q)$, while $\gamma^{[1]}$ does not change under
$\mathcal{L}$. The projection of $\mathbf{P}_{(1)}$ is
\begin{equation}
  \mathfrak{P}_{(1)}
  =\frac{1}{\vert
    q\vert}\left(q\cdot\frac{\bm{\partial}\>\>\>}{\bm{\partial} q} - 
  p\cdot\frac{\bm{\partial}\>\>\>}{\bm{\partial} p}\right),
\end{equation}
giving $\bm{\pi} = \bm{d}\left(q\cdot p\right)/\vert
q\vert$, and the primary constraint $\gamma^{[0]}=q\cdot p$.

The projection of $\overline{X}_L$ gives the Hamiltonian flow 
\begin{equation}
  \mathfrak{X}_C =
  \frac{\vert
    q\vert^2}{m}p\cdot\frac{\bm{\partial}\>\>\>}{\bm{\partial} q} -  
    \frac{\vert q\vert p^2}{m} \widehat{q}\cdot\frac{\bm{\partial}
      \>\>\>}{\bm{\partial} p} 
    - \frac{\partial V}{\partial
      q}\cdot\Pi\cdot\frac{\bm{\partial}\>\>\>}{\bm{\partial} p},
\end{equation}
and the total Hamiltonian $H_T = H_C +
u\gamma^{[0]}(\mathfrak{s})$. The projection of Eq.~$(\ref{stab})$ is 
\begin{equation}
  0= \frac{\vert q\vert^2}{m}p\cdot\frac{\partial \gamma^{[1]}}{\partial q} + 
  u\mathbf{U}^q_{(1)} \gamma^{[1]}. 
\end{equation}
For each of the three possible choices of $V(q)$ outlined above
the total Hamiltonian obtained here agrees with the one
obtained using constrained Hamiltonian mechanics.  

\subsection{A Lagrangian with Local
  Conformal Symmetry}

The action,
\begin{equation}
  S_{2} := \int\Bigg\{\frac{1}{2}m
  \left(\frac{d\widehat{q}_1}{dt}\right)^2+\frac{1}{2}m
  \left(\frac{d\widehat{q}_2}{dt}\right)^2+
  \frac{\lambda}{2}\left[\frac{q_1^a}{q_2}
  \frac{d\>\>}{dt}\left(\frac{q_{2a}}{\vert q_1\vert}\right) - \frac{q_2^a}{q_1}
  \frac{d\>\>}{dt}\left(\frac{q_{1a}}{\vert q_2\vert}\right)
  \right]
  \Bigg\} dt,
\end{equation}
where $a=1, \dots, d$, $D=2d$, describes an interacting, two particle
system that is invariant under the local conformal transformation
$q_1^a \to \alpha(\mathfrak{u}) q^a_1$ and $q_2^a \to
\alpha(\mathfrak{u}) q^a_2$. 

With
\begin{widetext}
\begin{eqnarray}
  \mathbf{\Omega}_M &=&  \frac{m}{\vert q_1\vert^2} \Pi_{ab}(q_1)
  \mathbf{d}q_1^a\wedge \mathbf{d} v_1^b + \frac{m}{\vert q_2\vert^2}
  \Pi_{ab}(q_2) \mathbf{d}q_2^a\wedge \mathbf{d} v_2^b, \hbox{  and}
  \nonumber
  \\
  \mathbf{\Omega}_F &=& \frac{m}{\vert q_1\vert^3}
  \left(\widehat{q}_1\cdot\mathbf{d}
  q_1\right)\wedge\left(v_1\cdot\Pi(q_1)\cdot \mathbf{d} q_1\right) + 
  \frac{m}{\vert q_2\vert^3} \left(\widehat{q}_2\cdot\mathbf{d} 
  q_2\right)\wedge\left(v_2\cdot\Pi(q_2)\cdot \mathbf{d} q_2\right)- 
  \nonumber
  \\
  &{}&
  \frac{\lambda}{\vert q_1\vert \vert
    q_2\vert}\left[\mathbf{d}q_1^a\wedge\left(\Pi(q_2)\cdot 
  \mathbf{d} q_2\right)_a+ \left(\Pi(q_1)\cdot
  \mathbf{d} q_1\right)_a\wedge \mathbf{d}q_2^a -\left(\Pi(q_1)\cdot
  \mathbf{d} q_1\right)^a\wedge \left(\Pi(q_2)\cdot
  \mathbf{d} q_2\right)_a\right]-
  \nonumber
  \\
  &{}& \frac{\lambda}{\vert q_1\vert^2}
  \left(\widehat{q}_1\cdot\mathbf{d}q_1\right)\wedge
  \left(\widehat{q}_2\cdot\Pi(q_1) \cdot \mathbf{d}q_1\right) +
  \frac{\lambda}{\vert q_2\vert^2} 
  \left(\widehat{q}_2\cdot\mathbf{d}q_2\right)\wedge
  \left(\widehat{q}_1\cdot\Pi(q_2) \cdot \mathbf{d}q_2\right),
\end{eqnarray}
\end{widetext}
$\mathcal{C}$ and $\mathcal{G}$ are two-dimensional and
are spanned by
\begin{equation}
  \mathbf{U}^q_{(1)} =
  \widehat{q}_1\cdot\frac{\bm{\partial}\>\>\>}{\bm{\partial} q_1},
    \quad
  \mathbf{U}^q_{(2)} =
  \widehat{q}_2\cdot\frac{\bm{\partial}\>\>\>}{\bm{\partial} q_2},
  \quad \hbox{and}\quad 
  \mathbf{U}^v_{(1)} =
  \widehat{q}_1\cdot\frac{\bm{\partial}\>\>\>}{\bm{\partial} v_1},
    \quad
  \mathbf{U}^v_{(2)} =
  \widehat{q}_2\cdot\frac{\bm{\partial}\>\>\>}{\bm{\partial} v_2},
\end{equation}
respectively. The reduced $\bar{F}=0$, and $\overline{\hbox{ker
  }\mathbf{\Omega}_L(\mathfrak{u})}$ is spanned by $\mathbf{U}^v_{(1)},
\mathbf{U}^v_{(2)}$, and 
\begin{eqnarray}
  \mathbf{P}_{(1,2)} &=&
  \widehat{q}_{1,2}\cdot\frac{\bm{\partial}\>\>\>}{\bm{\partial} q_{1,2}}
  +
  \frac{v_{1,2}}{\vert
    q_{1,2}\vert}\cdot\Pi(q_{1,2})\cdot\frac{\bm{\partial}\>\>\>}{\bm{\partial}v_{1,2}} 
  +
  \nonumber
  \\
  &{}&(-1)^{1,2}
  \frac{\lambda}{m}
  \left[\widehat{q}_{2,1}\cdot\Pi(q_{1,2})\cdot\frac{\bm{\partial}\>\>\>}{\bm{\partial}
      v_{1,2}}
    +\frac{\vert q_{2,1}\vert}{\vert q_{1,2}\vert}\widehat{q}_{1,2}\cdot\Pi(q_{2,1})\cdot
    \frac{\bm{\partial}\>\>\>}{\bm{\partial} v_{2,1}}\right],
\end{eqnarray}
The energy is
\begin{equation}
  E=\frac{1}{2}\frac{m}{\vert q_1\vert^2}v_1\cdot\Pi(q_1)\cdot v_1 +
  \frac{1}{2}\frac{m}{\vert q_2\vert^2}v_2\cdot\Pi(q_2)\cdot v_2.
  \label{energy2}
\end{equation}
Although $\mathcal{C}$ is two-dimensional, 
$\gamma^{[1]}_{(1)} = -\lambda \gamma^{[1]}/\vert q_1\vert$ and
$\gamma^{[1]}_{(2)} = \lambda \gamma^{[1]}/\vert q_2\vert$, and the two
first-order Lagrangian constraints reduce to one
\begin{equation}
  \gamma^{[1]} = \widehat{q}_2\cdot \Pi(q_1)\cdot \frac{v_1}{\vert q_1\vert} +
  \widehat{q}_1\cdot \Pi(q_2) \cdot \frac{v_2}{\vert q_2\vert},
\end{equation}
with
\begin{eqnarray}
  \bm{\beta}[\mathbf{X}_E] &=& -\lambda\gamma^{[1]}
  \left(\frac{\mathbf{\Theta}_q^{(1)}}{\vert q_1\vert} -
  \frac{\mathbf{\Theta}_q^{(2)}}{\vert q_2\vert}\right).
\end{eqnarray}
As expected, $\mathbf{G}\gamma^{[1]}=0$ for any
$\mathbf{G}\in\mathcal{G}$. 
We may choose 
  \begin{eqnarray}
  \mathbf{X}_{L}&=&v_1\cdot\frac{\bm{\partial}\>\>\>}{\bm{\partial}
    q_1} +v_1\cdot\frac{\bm{\partial}\>\>\>}{\bm{\partial}
    q_1} + \left[2\frac{(\widehat{q}_1\cdot v_1)}{\vert q_1\vert} v_1
    +\frac{\lambda}{m} \left(\frac{\vert q_1\vert}{\vert q_2\vert} v_2
    - (\widehat{q}_1\cdot 
    v_1) \widehat{q}_2\right)\right]\cdot
  \frac{\bm{\partial}\>\>\>}{\bm{\partial} v_1}+
  \nonumber
  \\
  &{}&
  \left[2\frac{(\widehat{q}_2\cdot v_2)}{\vert q_2\vert} v_2
    -\frac{\lambda}{m} \left(\frac{\vert q_2\vert}{\vert q_1\vert} v_1 - (\widehat{q}_2\cdot
    v_2) \widehat{q}_1\right)\right]\cdot
  \frac{\bm{\partial}\>\>\>}{\bm{\partial} v_2}.
\end{eqnarray}
As $[\mathbf{X}_{L},\mathbf{U}^v_{(1,2)}] \sim - \mathbf{P}_{(1,2)}/q_1$, 
the action on $\mathbf{X}_L$ by
$\hbox{Gr}_{\bm{\mathfrak{S}}\hbox{\textbf{ym}}}$ does not give a
SOLVF. Instead, we construct  
\begin{eqnarray}
\overline{\mathbf{X}}_{L} &=&
v_1\cdot\Pi(q_1)\cdot\frac{\bm{\partial}\>\>\>}{\bm{\partial}
  q_1} + v_2\cdot\Pi(q_2)\cdot\frac{\bm{\partial}\>\>\>}{\bm{\partial}
  q_2} +
\nonumber
\\
&{}&
\left(\frac{\widehat{q}_1\cdot v_1}{\vert q_1\vert}\right)
v_1\cdot\Pi(q_1)\cdot\frac{\bm{\partial}\>\>\>}{\bm{\partial} v_1} +
\left(\frac{\widehat{q}_2\cdot v_2}{\vert q_2\vert }\right)
v_2\cdot\Pi(q_2)\cdot\frac{\bm{\partial}\>\>\>}{\bm{\partial} v_2} 
+
\nonumber
\\
&{}&
\frac{\lambda}{m} \left(\frac{\vert q_1\vert}{\vert q_2\vert}
  v_2\cdot\Pi(q_2)\cdot\Pi(q_1)\cdot \frac{\bm{\partial}
    \>\>\>}{\bm{\partial} v_1}-\frac{\vert q_2\vert}{\vert q_1\vert}
  v_1\cdot\Pi(q_1)\cdot\Pi(q_2)\cdot \frac{\bm{\partial}
    \>\>\>}{\bm{\partial} v_2} \right).
\end{eqnarray}
A general SOELVF is then $\overline{\mathbf{X}}_{EL} = \overline{\mathbf{X}}_{L} +
  u^{(-)}(\mathfrak{u})\left[\mathbf{P}_{(-)}\right] +
  u^{(+)}(\mathfrak{u}) \left[\mathbf{P}_{(+)}\right]$, 
where $u^{(\pm)}(\mathfrak{u})\in\overline{\mathcal{F}}$ and
$\mathbf{P}_{(\pm)} = \vert q_1\vert\mathbf{P}_{(1)} \pm
\vert q_2\vert\mathbf{P}_{(2)}$. One of the arbitrary functions
\begin{equation}
  u^{(-)}(\mathfrak{u}) = \frac{m}{4\lambda}\frac{i_{\overline{\mathbf{X}}_{L}}\mathbf{d}\gamma^{[1]}}{ 
    \left[1-(\widehat{q}_1\cdot\widehat{q}_2)\right]},
  \label{u2}
\end{equation}
is determined through the constraint algorithm with
\begin{eqnarray}
  i_{\overline{\mathbf{X}}_{L}}\mathbf{d}\gamma^{[1]}&=& -
  2(\widehat{q}_1\cdot\widehat{q}_2)\frac{E}{m}+ \frac{2}{\vert
    q_1\vert \vert q_2\vert}
  v_1\cdot\Pi(q_1)\cdot\Pi(q_2)\cdot v_2
  -
  \nonumber
  \\
  &{}&
  \frac{\lambda}{m}(\widehat{q}_1\cdot\widehat{q}_2)\left[v_2\cdot\Pi(q_2)\cdot\widehat{q}_1
    - v_1\cdot\Pi(q_1)\cdot \widehat{q}_2\right].
\label{g2}
\end{eqnarray}
The other one, $u^{(+)}(\mathfrak{u})$, is not. 

We find that $\mathfrak{L}_{\mathbf{P}_{(+)}}
\bm{\beta} = 0$, while
\begin{equation}
  \mathfrak{L}_{\mathbf{P}_{(-)}} \bm{\beta} =
  -\frac{4\lambda}{m}\left[1-(\widehat{q}_1\cdot\widehat{q}_2)^2\right].
\end{equation}
Then $\mathcal{S}\hbox{ym}$ is one-dimensional, and
spanned by $\mathbf{P}_{(+)}$.

With the canonical momenta,
\begin{equation}
  p_{1a} := \frac{m}{\vert q_1\vert^2} \Pi_{ab}(q_1)v^b_1 -
  \frac{\lambda}{2}\frac{\tau_{ab}(q_1)}{\vert q_1\vert}\widehat{q}^b_2, \quad
  p_{2a} := \frac{m}{\vert q_2\vert^2} \Pi_{ab}(q_2)v^b_2 +
  \frac{\lambda}{2}\frac{\tau_{ab}(q_2)}{\vert q_2\vert}\widehat{q}^b_1,
  \label{mom}
\end{equation}
where $\tau_{ab}:= \delta_{ab}+\widehat{q}_a\widehat{q}_b$, the passage
to Hamiltonian mechanics is straightforward. Equation $(\ref{energy2})$
gives $H_C = \vert q_1\vert^2L_1^2/2m + \vert q_2\vert^2L_2^2/2m$ with 
\begin{equation}
  L_{1a}:=\vert q_1\vert p_{1a} +
    \frac{\lambda}{2}\tau_{ac}(q_1)\widehat{q}_2^c, \quad 
    L_{2a}:= \vert q_2\vert p_{2a} -
\frac{\lambda}{2}\tau_{ac}(q_2)\widehat{q}_1^c,
\end{equation}
and the projection of the first-order Lagrangian constraint is $\gamma^{[1]}
=\left[\widehat{q}_1\cdot L_2  +  \widehat{q}_2\cdot
  L_1\right]/m$. The projection of $\mathbf{P}_{(\pm)}$ is
\begin{widetext}
  \begin{eqnarray}
  \mathfrak{P}_{(+)} &=&
    \widehat{q_1}\cdot\frac{\bm{\partial}\>\>\>}{\bm{\partial}q_1} +
  \widehat{q_2}\cdot\frac{\bm{\partial}\>\>\>}{\bm{\partial}q_2} 
  -p_1\cdot\frac{\bm{\partial}\>\>\>}{\bm{\partial} p_1}
    -p_2\cdot\frac{\bm{\partial}\>\>\>}{\bm{\partial} p_2},
      \nonumber
      \\
  \mathfrak{P}_{(-)} &=&
  \widehat{q_1}\cdot\frac{\bm{\partial}\>\>\>}{\bm{\partial}q_1} -
  \widehat{q_2}\cdot\frac{\bm{\partial}\>\>\>}{\bm{\partial}q_2} 
  -p_1\cdot\frac{\bm{\partial}\>\>\>}{\bm{\partial} p_1}
  +p_2\cdot\frac{\bm{\partial}\>\>\>}{\bm{\partial} p_2}-
 \nonumber
  \\
  &{}&
  2\lambda\left[
    \frac{\widehat{q}_2}{\vert q_1\vert}\cdot\Pi(q_1)\cdot\frac{\bm{\partial}\>\>\>}{\bm{\partial}
      p_1} + \frac{\widehat{q}_1}{\vert q_2\vert}\cdot\Pi(q_2)\cdot\frac{\bm{\partial}\>\>\>}{\bm{\partial}
      p_2}\right],
\end{eqnarray}
\end{widetext}
giving $\bm{\pi}_{(+)} = \bm{d}\left(q_1\cdot p_1+ q_2\cdot
p_2\right)$, and $\bm{\pi}_{(-)} = \bm{d}\left(q_1\cdot p_1 - q_2\cdot
p_2 +  \lambda \widehat{q}_1\cdot\widehat{q}_2\right)$. 
The primary Hamiltonian constraints are $\gamma^{[0]}_{(+)}:= 
q_1\cdot p_1+ q_2\cdot p_2$ and $\gamma^{[0]}_{(-)} := q_1\cdot p_1 -
q_2\cdot p_2 + \lambda \widehat{q}_1\cdot\widehat{q}_2$.

The projection of $\overline{\mathbf{X}}_L$ gives
\begin{widetext}
  \begin{eqnarray}
  \mathfrak{X}_{H_C} &=& \frac{\vert q_1\vert}{m}
  L_1\cdot\frac{\bm{\partial}\>\>\>}{\bm{\partial} q_1} + \frac{\vert
    q_2\vert}{m} L_2\cdot\frac{\bm{\partial}\>\>\>}{\bm{\partial} q_2} -
  \nonumber
  \\
  &{}&
\frac{\lambda}{2m\vert q_1\vert}
  \left[\left(\widehat{q}_1\cdot\widehat{q}_2\right)L_1
    -L_2\right]\cdot\frac{\bm{\partial}\>\>\>}{\bm{\partial}p_1}
  -\frac{\left[L_1^2+\lambda\left(\widehat{q}_1\cdot
    L_2\right)\right]}{m\vert q_1\vert}
  \widehat{q}_1\cdot\frac{\bm{\partial}\>\>\>}{\bm{\partial}p_1}+ 
  \nonumber
  \\
  &{}&
  \frac{\lambda}{2m\vert q_2\vert}
  \left[\left(\widehat{q}_1\cdot\widehat{q}_2\right)L_2
    -L_1\right]\cdot\frac{\bm{\partial}\>\>\>}{\bm{\partial}p_2}
  -\frac{\left[L_2^2+\lambda\left(\widehat{q}_2\cdot
    L_1\right)\right]}{m\vert q_2\vert}
  \widehat{q}_2\cdot\frac{\bm{\partial}\>\>\>}{\bm{\partial}p_2}. 
\end{eqnarray}
\end{widetext}
Then, $H_T=H_C+u^{(-)}\gamma^{[0]}_{(-)}(\mathfrak{s})
+u^{(+)}\gamma^{[0]}_{(+)}(\mathfrak{s})$, 
where after using Eq.~$(\ref{mom})$ in Eqs.~$(\ref{u2})$ and
$(\ref{g2})$, 
\begin{equation}
  u^{(-)} = \frac{1}{2\lambda}\left[
  \frac{
    L_1\cdot L_2/m + \lambda(\widehat{q}_1\cdot\widehat{q}_2)\widehat{q}_1\cdot L_2
    -(\widehat{q}_1\cdot\widehat{q}_2)H_C
  }{
    1-(\widehat{q}_1\cdot\widehat{q}_2)^2}
  \right],
\end{equation}
while $u^{(+)}$ remains undetermined.

\subsection{A Lagrangian with Local Conformal and
  Time-reparametization Invariance}

The action
\begin{equation}
  S_{3} := sm\int\left[s\left(\frac{d\widehat{q}}{dt}\right)^2\right]^{1/2} dt,
\end{equation}
where $s=\pm1$, is invariant under both the local conformal
transformations, $q^a\to \alpha(\mathfrak{u}) q^a$, and the
reparametization $t\to \tau(t)$, where $\tau$ is a
monotonically increasing function of $t$ (see also \cite{Jam2019} and
\cite{Rab2003, Mus2005} for systems with Lagrangians linear in the
velocities). This action is a generalization of that for the
relativistic particle, with the additional requirement that it have a 
local conformal invariance.   

For this action,
\begin{equation}
  \mathbf{\Omega}_L = \frac{m}{\vert q\vert}
  \frac{P_{ab}(u)}{\sqrt{sv\cdot \Pi(q)}\cdot
      v}\mathbf{d}q^a\wedge\mathbf{d}v^b,
\end{equation}
and $\mathbf{\Omega}_F=0$. Here, $a=1. \dots, D$,
\begin{equation}
  u_a=\frac{\Pi_{ab}(q)v^b}{\sqrt{sv\cdot\Pi(q)\cdot v}},
\end{equation}
so that $u^2 = s$, while $P_{ab}(u) = \Pi_{ab}(q) -su_a u_b$. Then ker
$\mathbf{\Omega}_L(\mathfrak{u}) = $ ker 
$\mathbf{\Omega}_M(\mathfrak{u})$, and both $\mathcal{C}$ and
$\mathcal{G}$ are two-dimensional. They are spanned by
\begin{equation}
  \mathbf{U}^q_{(1)} = 
  \widehat{q}\cdot\frac{\bm{\partial}\>\>\>}{\bm{\partial} q}, \quad
    \mathbf{U}^q_{(2)} =
  u\cdot\frac{\bm{\partial}\>\>\>}{\bm{\partial} q}, \quad
  \hbox{and}\quad 
    \mathbf{U}^v_{(1)} =
  \widehat{q}\cdot\frac{\bm{\partial}\>\>\>}{\bm{\partial} v}, \quad
    \mathbf{U}^v_{(2)} =
    u\cdot\frac{\bm{\partial}\>\>\>}{\bm{\partial} v},
\end{equation}
respectively. 

Because this system is fully constrained, $E=0$. As
$\mathbf{\Omega}_F=0$ as well, there are no Lagrangian constraints. We
may choose $\mathbf{X}_{L} = 
v\cdot\bm{\partial}/\bm{\partial}q$. As $[\mathbf{X}_{L},
  \mathbf{U}^v_{(1,2)}] \sim-\mathbf{U}^q_{(1,2)}$, the action of
$\mathbf{X}_L$ by $\hbox{Gr}_{\mathcal{S}\hbox{ym}}$ does
not give a SOLVF. The vector field $\overline{\mathbf{X}}_L$ can be
constructed, and as expected for a fully constrained system,
$\overline{\mathbf{X}}_{L} = 0$. A general SOELVF is then
$\overline{\mathbf{X}}_{EL} = u^{1}(\mathfrak{u}) 
\left[\mathbf{U}^q_{(1)}\right] + u^{2}(\mathfrak{u})
\left[\mathbf{U}^q_{(2)}\right]$, with
$u^{n}(\mathfrak{u})\in\overline{\mathcal{F}}$ for $n=1,2$. The Lie
algebra $\mathcal{S}\hbox{ym}$ itself is two dimensional, and spanned by 
$\mathbf{U}^q_{(1)}$ and $\mathbf{U}^q_{(2)}$.   

For the passage to Hamiltonian mechanics, $H_C=0$ as $E=0$. With the
canonical momentum $p_a = mu_a/\vert q\vert$, 
the projection of $\mathbf{P}_{(1,2)}$ is 
\begin{equation}
  \mathfrak{P}_{(1)} = \widehat{q}\cdot\frac{\bm{\partial}\>\>\>}{\bm{\partial}
    q} - \frac{1}{\vert
    q\vert}p\cdot\frac{\bm{\partial}\>\>\>}{\bm{\partial} p}, \>\> 
  \mathfrak{P}_{(2)} =
  \frac{q}{m}p\cdot\frac{\bm{\partial}\>\>\>}{\bm{\partial} q} - 
  \frac{p^2}{m}\widehat{\vert q\vert}\cdot\frac{\bm{\partial}
    \>\>\>}{\bm{\partial} p}, 
\end{equation}
and $\bm{\pi}_{(1)} = \bm{d}(q\cdot p)/\vert q\vert,
\>\> \bm{\pi}_{(2)} = \bm{d}(\vert q\vert^2p^2)/2m\vert q \vert$. The
primary Hamiltonian constraints are $\gamma^{[0]}_1 :=  
q\cdot p$ and $\gamma^{[0]}_2 := \vert q\vert^2p^2-sm^2$. As
$\overline{\mathbf{X}}_L=0$, $\mathfrak{X}_C=0$, and  
we find that
\begin{equation}
  H_T = \frac{u_{(1)}}{\vert q\vert} \gamma^{[0]}_1(\mathfrak{s}) +
  \frac{u_{(2)}}{2m\vert q\vert}\gamma^{[0]}_2(\mathfrak{s}).
\end{equation}

\section{Concluding Remarks\label{Conc}}

With the benefit of hindsight, the many
roles that $\mathcal{G}$ plays in determining both the geometric
structure of $\mathbb{P}_L$ for singular Lagrangians, and the
connection between these structures and dynamics become readily
apparent. What also becomes clear are the reasons why SOELVFs and
their dynamical structures are projectable.

Because $\mathcal{G}$ is involutive, it gives a foliation of
$\mathbb{P}_L$. There is then a neighborhood $U$ about each point
$\mathfrak{u}\in\mathbb{P}_L$ on which we can define the equivalence
relation $\mathfrak{u}_1\sim\mathfrak{u}_2$ iff $\mathfrak{u}_1-\mathfrak{u}_2
=g$, where $g$ is a point on the leaves
${\mathfrak{F}\hbox{\textbf{ol}}}(\mathcal{G})$ of the
foliation. This leads to the quotient space
$\mathbb{P}_L/{\mathfrak{F}\hbox{\textbf{ol}}}(\mathcal{G})$, which has
dimension $2D-N_0$. Importantly, 
$\mathbb{P}_L/{\mathfrak{F}\hbox{\textbf{ol}}}(\mathcal{G})$ is projectable, 
and $\mathcal{L}(\mathbb{P}_L/{\mathfrak{F}\hbox{\textbf{ol}}}(\mathcal{G}))
=\mathcal{L}(\mathbb{P}_L)= \mathbb{P}_C^{[0]}$. This structure, and
the role that $\mathcal{G}$ plays in its construction, is
well known in the literature \cite{Got1979, Car1988a}.

Next, because
$\mathcal{G}\subset \mathbf{T}_{\mathfrak{u}}\mathbb{P}_L$ and
is involutive, it is natural to follow the construction of
$\mathbb{P}_L/{\mathfrak{F}\hbox{\textbf{ol}}}(\mathcal{G})$
and consider the 
set of vector fields in $\mathbf{T}_{\mathfrak{u}}\mathbb{P}_L$ for
which $\mathcal{G}$ is an ideal. This leads us to
$\overline{\mathbf{T}_{\mathfrak{u}}\mathbb{P}_L}$, and the quotient
space
$\overline{\mathbf{T}_{\mathfrak{u}}\mathbb{P}_L}/\mathcal{G}$. As
dim
$\overline{\mathbf{T}_{\mathfrak{u}}\mathbb{P}_L}/\mathcal{G}=2D-N_0$,
it is expected that  
$\overline{\mathbf{T}_{\mathfrak{u}}\mathbb{P}_L}/\mathcal{G}=\mathbf{T}_{\mathfrak{p}}\left[\mathbb{P}_L/{\mathfrak{F}\hbox{\textbf{ol}}}(\mathcal{G})\right]$   
for
$\mathfrak{p}\in\mathbb{P}_L/{\mathfrak{F}\hbox{\textbf{ol}}}(\mathcal{G})$. That   
$\overline{\mathbf{T}_{\mathfrak{u}}\mathbb{P}_L}/\mathcal{G}$
is projectable is then readily apparent. 

Finally, for
singular Lagrangians the acceleration is  
not determined uniquely by the Euler-Lagrange equations of
motion, an ambiguity due to the generalized Lie symmetry. This symmetry
is generated by vectors that must lie in the kernel of
$\bm{\Omega}_L$, and yet cannot be in
$\mathcal{G}$, leading naturally first to the construction of 
ker $\overline{\mathbf{\Omega}(\mathfrak{u})}/\mathcal{G}$,
and then to the construction of $\overline{\mathcal{S}\hbox{ol}}$. Both
ker $\overline{\mathbf{\Omega}(\mathfrak{u})}/\mathcal{G} \subset
\overline{\mathbf{T}_{\mathfrak{u}}\mathbb{P}_L}/\mathcal{G}$ and 
$\overline{\mathcal{S}\hbox{ol}}\subset
\overline{\mathbf{T}_{\mathfrak{u}}\mathbb{P}_L}/\mathcal{G}$, and thus the
evolution of the dynamical system is confined to the tangent bundle
$\mathbf{T}\left[\mathbb{P}_L/{\mathfrak{F}\hbox{\textbf{ol}}}(\mathcal{G})\right]$. The projectability of 
$\mathbf{T}[\mathbb{P}_L/{\mathfrak{F}\hbox{\textbf{ol}}}(\mathcal{G})]$ 
ensures that all of the dynamical structures needed to describe the
evolution of dynamical systems on the Lagrangian phase space is
projectable, and agrees with those
obtained through constrained Hamiltonian mechanics. 

While $\mathcal{G}$ does play an important role in determining the
general Lie symmetry group, it itself is not the generator of this
group. This can be readily seen in the first example in
\textbf{Section \ref{Exam}} where the Lagrangian may or may not have a
local gauge symmetry depending on the choice of potential. Nevertheless,
$\mathcal{G}$ is present and plays its usual role in determining 
$\mathbb{P}_L/\mathfrak{F}\hbox{\textbf{ol}}(\mathcal{G})$. It
is instead vectors in $\overline{\hbox{ker
  }\mathbf{\Omega}_L(\mathfrak{u})}/\mathcal{G}$\textemdash with
$\mathcal{G}$ removed\textemdash that generate the
generalized Lie symmetry. We emphasize here that while this symmetry plays
an important and guiding role, this role is nevertheless supportive in
the construction of the algebraic-geometric structures on
$\mathbb{P}_L$ needed in determining both the geometric structure of
$\mathbb{P}_L$, and the connection between these structures and the
evolution of the dynamical system. 

The application of these algebraic-geometric structures go beyond
showing the equivalence of the Lagrangian and Hamiltonian formulations
of mechanics for singular Lagrangians, however. While the primary Hamiltonian
constraints play a critical role in the Hamiltonian constraint
analysis, the constraints themselves have traditionally been found by
inspection; the expectation is that this inspection is able to both
determine their form and to ensure that all of the constraints has been
found for the system at hand. As
a result of the Lagrangian phase space analysis presented here we are
able to determine the number of primary constraints for any dynamical
system, and the constraints themselves can be calculated by solving a 
first-order, quasi-linear differential equation. In addition, while
the end result of the Lagrangian constraint algorithm is a SOELVF
defined in terms of a certain number of arbitrary functions, and the
end result of the Hamiltonian constraint analysis is a total
Hamiltonian with the same number of arbitrary functions, how many
arbitrary functions are needed, and their relationship to the original
symmetries of the action is not known. With direct access to the
Lagrangian and its symmetries, these questions can now be addressed in
the Lagrangian phase space formulation.

\begin{acknowledgments}
  This paper would not have been possible without the contributions by
  John Garrison, who provided most of the essential mathematics in
  \textbf{Section \ref{&LagnPhaseSp}}. Publication made possible in
  part by support from the Berkeley Research Impact Initiative (BRII)
  sponsored by the UC Berkeley Library. 
\end{acknowledgments}


\begin{thebibliography}{60}
\bibitem{foot0}
  There are two definitions of `presymplectic'\textemdash a broader one
  in which $\mathbf{\Omega}_L(\mathfrak{u})$ need only be closed, and
  a more narrow one that also requires the rank of
  $\mathbf{\Omega}_L(\mathfrak{u})$ to be constant on
  $\mathbb{P}_L$\textemdash used in the literature. Although in this
  paper our focus is on systems where the rank
  $\mathbf{\Omega}_L(\mathfrak{u})$ is constant, a few of our
  results still hold even when it is not, and we follow the broader
  definition of the word.
\bibitem{foot3}
    We use $\mathfrak{u}$ to represent both $(q,\dot{q})$ for trajectories
    on $\mathbb{Q}$ and $(q,v)$ for trajectories on $\mathbb{P}_L$ to
    avoid introducing additional notation.
\end{thebibliography}
\end{document}